\documentclass[pra,twocolumn,showpacs,preprintnumbers,amsmath,amssymb,superscriptaddress,longbibliography]{revtex4-2}
\usepackage{graphicx}
\usepackage{dcolumn}
\usepackage{bm}
\usepackage{epsfig}
\usepackage{amsmath}
\usepackage{amssymb}
\usepackage{empheq}
\usepackage{color}
\usepackage{bbm}
\usepackage[caption=false]{subfig}
\usepackage{placeins}
\usepackage{soul}
\usepackage{physics}
\usepackage{mathrsfs}
\usepackage{url}
\usepackage{changes} 
\definechangesauthor[name={Per cusse}, color=red]{per} 
\usepackage[colorlinks=true,citecolor=blue,linkcolor=blue,urlcolor=blue]{hyperref}
\begin{document}
	\title{Analytical approach to higher-order correlation functions \\in U(1) symmetric systems}
	\author{Zhi-Guang Lu}
	\affiliation{School of Physics, Huazhong University of Science and Technology, Wuhan, 430074, P. R. China}
	
	\author{Cheng Shang}
	\affiliation{Department of Physics, The University of Tokyo, 5-1-5 Kashiwanoha, Kashiwa, Chiba 277-8574, Japan}
	
	\author{Ying Wu}
	\affiliation{School of Physics, Huazhong University of Science and Technology, Wuhan, 430074, P. R. China}
	
	\author{Xin-You L\"{u}}\email{xinyoulu@hust.edu.cn}
	\affiliation{School of Physics, Huazhong University of Science and Technology, Wuhan, 430074, P. R. China}
	
	\date{\today}
	\begin{abstract}
		We derive a compact analytical solution of the $n$th-order equal-time correlation functions by using scattering
		matrix ($S$ matrix) under a weak coherent state input. Our solution applies to any dissipative quantum system that
		respects the U(1) symmetry. We further extend our analytical solution into two categories depending on whether
		the input and output channels are identical. The first category provides a different path for studying cross-correlation
		and multiple-drive cases, while the second category is instrumental in studying waveguide quantum
		electrodynamics systems. Our analytical solution allows for easy investigation of the statistical properties of
		multiple photons even in complex systems. Furthermore, we have developed a user-friendly open-source library
		in Python known as the quantum correlation solver, and this tool provides a convenient means to study various dissipative quantum systems that satisfy the above-mentioned criteria. Our study enables using $S$ matrix to study
		the photonic correlation and advance the possibilities for exploring complex systems.
	\end{abstract}
	
	\maketitle
	
	\section{Introduction}
	The correlation function is crucial in various fields, including condensed matter physics \cite{Peskin1995AnIT}, statistical physics \cite{Pathria1996}, and quantum physics \cite{scully_zubairy_1997}. Correlation functions are particularly in quantum physics for characterizing the statistical properties of light. Specifically, they play a vital role in developing various quantum devices, such as scalable coherent single-photon source devices \cite{PhysRevLett.123.250503,single_photon1,shang2023coupling}, nonreciprocal quantum devices \cite{PhysRevX.5.021025,PhysRevApplied.7.024028,PhysRevX.7.031001,PhysRevLett.120.060601,Shang:19}, and two-photon devices \cite{PhysRevLett.118.133604}, which are essential for quantum information processing \cite{bennett_quantum_2000,Buluta_2011} and quantum engineering \cite{PhysRevLett.81.3611}. 
	
	The second-order equal-time correlation function (ETCF) is the most straightforward and typical example used to describe the statistical properties of light. The value of the correlation function characterizes the photon-number statistics of light. For instance, when the correlation function is less than one, it describes the sub-Poissonian photon-number statistics \cite{PhysRevA.41.475, PhysRevLett.108.183601}. While it is more than one, it characterizes the super-Poissonian photon-number statistics \cite{scully_zubairy_1997}. Similarly, higher-order ETCFs can be utilized to reveal specific effects like photon-induced tunneling and multi-photon blockade \cite{PhysRevLett.121.153601}. 
	
	However, despite being simple open quantum systems with nonlinearities, such as a two-level atom trapped in a cavity or an optical resonator with Kerr-type nonlinearity, the analytical computation of higher-order ETCFs remains challenging. So far, several effective methods have been proposed to address this challenge, such as the master-equation method \cite{lindblad_generators_1976,Gorini1976CompletelyPD,open_system} and quantum-trajectory approach \cite{RevModPhys.70.101,Qt}. The common feature of the above methods is that $n$th-order ETCF is treated by numerical calculation. Notably, the analytical solution is more intrinsic, and scattering matrix (S-matrix) methods \cite{SM10,SM12,SM15,SM9,SM8,SM14,SM11,SM1,SM3,SM7,SM5,SM6,SM4,SM2,SM13} allow us to resolve the problem by treating the low-dimensional system, such as quantum dots, superconducting qubits, and atomic ensembles, as a scatterer or potential field for the optical fields and attempting to relate the incoming and outgoing optical fields \cite{PhysRevB.98.144112}.
	
	In this paper, we theoretically demonstrate that for a large class of open quantum systems satisfying the U(1) symmetry under a weak coherent state input, the $n$th-order ETCF could be described entirely by the $n$-photon $S$ matrix. Subsequently, we give the $n$-photon $S$ matrix for arbitrary input and output channels, and the $S$ matrix completely depends on Green's function. In order to further calculate the $S$ matrix, we prove that the time-ordered Green's function can be exactly computed by using an effective Hamiltonian that involves only the degrees of freedom of the system without the part of environments which has an infinite number of degrees of freedom. To proceed, we focus on studying a system that respects the conservation of the total excitation number. Therefore, the effective Hamiltonian could be decomposed into a block-diagonal form, and the annihilation operators and creation operators of the system have a similar matrix form, i.e., block-upper and -lower triangular forms. As conclusions above, we first define a probability amplitude of equal-time probing multiple photons and give the concrete expression. Therefore, we derive a compact analytical solution of the $n$th-order ETCF, and the analytical expression ultimately depends on the probability amplitude we defined. Moreover, the compact analytical solution can be extended from the single-mode coherent drive to the multi-mode coherent drive and from the different input and output channels to the same. As a result, the former provides a different way to study the dynamical photon blockade phenomenon \cite{dyna_pb1,dyna_pb2}, and the latter opens up a different path to explore the waveguide quantum electrodynamics system \cite{Shen:05}.
	
	The paper is organized as follows: In Sec.~\ref{II}, we introduce a total Hamiltonian including system and environments, and then prove the equivalence of two types of $n$th-order correlation functions. In Sec.~\ref{III}, we derive the $n$-photon $S$ matrix and the effective Hamiltonian. In Sec.~\ref{IV}, we define a probability amplitude of equal-time probing multiple photons in order to acquire the compact analytical solution of $n$th-order ETCF. In Sec.~\ref{V}, we study the effect of two categories corresponding to different and identical input-output channels, respectively. The former further divides into four cases, i.e., one-to-one [In Sec.~\ref{VA}], many-to-one [In Sec.~\ref{VB}], one-to-many [In Sec.~\ref{VC}] and many-to-many. The latter is described by Sec.~\ref{VD}. Finally, in Sec.~\ref{VI}, we calculate three examples and give the concise analytical solution of the first example as a representative case for illustrating the versatility and validity of the analytical solution.
	
	\section{Equivalent Model}\label{II}
	We first consider a generalized open quantum system satisfying the conservation of total excitation number, which means that the U(1) symmetry is satisfied. For the sake of simplicity, the system Hamiltonian on which we focused is represented by $H_{\text{sys}}\{o_k\}$, where the notation $\{o_k\}$ indicates that the system of interest comprises several local system modes. Subsequently, we introduce the symbol $\mathcal{N}$ to represent the total excitation number operator of the system, and the operator commutes with the system Hamiltonian, i.e., $[H_{\text{sys}}, \mathcal{N}]=0$. Finally, without loss of generality, we assume that each local system interacts with one or more individual heat baths (defaulting to two), and the interaction Hamiltonian $H_{\text{I}}$ between the system and the heat baths does not break the U(1) symmetry. Consequently, the total Hamiltonian $H_{\text{tot}}$, involving both the system and the heat baths, is defined as ($\hbar=1$)\,\cite{PhysRevA.31.3761}
	\begin{align}
		H_{\text{tot}}&=H_{\text{sys}}\{o_k\} + H_{\text{B}}+H_{\text{I}}\quad\rm{with}\label{eq1}\\
		H_{\text{B}}&=\int \dd\omega \sum_k\omega\left[b_k^\dagger(\omega) b_k(\omega)+c_k^\dagger(\omega) c_k(\omega)\right]\nonumber,\\
		H_{\text{I}}&=\int\dd\omega\sum_k\left[\xi_{b,k}b_k^\dagger(\omega)o_k+\xi_{c,k}c_k^\dagger(\omega) o_k+\text{H.c.}\right]\nonumber,
	\end{align}
	where $\xi_{b,k}$ ($\xi_{c,k}$) denotes the coupling strength  between the heat bath with mode $b_k(\omega)$ [$c_k(\omega)$] and the $k$-th local system, and is assumed to be frequency independent. $o_k$ represents a lowering operator of the $k$-th local system that is assumed to commute with the heat bath modes, e.g., $b_k(\omega)$ and $b_k^\dagger(\omega)$. In this section we assume $o_k$ to be arbitrary. In practice, $o_k$ can be a bosonic annihilation operator describing a cavity mode or a lowering operator for the two-level atom. $b_k(\omega)$ [$b_k^\dagger(\omega)$] and $c_k(\omega)$ [$c_k^\dagger(\omega)$] both are bosonic annihilation (creation) operators of the heat baths coupled to the $k$-th local system. Note that the subscript $k$ only is used to label the local system and the corresponding heat baths, and in some specific models we also do not necessarily introduce the subscript. Meanwhile, the operators also satisfy the standard commutation relation:
	\begin{align}\label{eq2}
		[\mu_m(\omega), \nu_l(\omega^\prime)]=\delta_{m,l}\delta_{\mu,\nu}\delta(\omega-\omega^\prime)\quad  \mu,\nu\in\{b,c\}.
	\end{align}
	Notice that the Hamiltonian $H_{\text{I}}$ in Eq.~(\ref{eq1}) is a linear interaction. For the case of a class of nonlinear interactions satisfying the U(1) symmetry, the specific discussions are presented in Appendix \ref{E}.
	
	To proceed, we introduce a laser coherently driving to the $i$-th local system described by Eq.~(\ref{eq1}), and the corresponding Hamiltonian is given by
	\begin{eqnarray}
		\begin{aligned}\label{eq3}
			H_{\text{d}}= (\Omega_i^*o_ie^{i\omega_dt}+\Omega_io_i^\dagger e^{-i\omega_dt}),
		\end{aligned}
	\end{eqnarray}
	where $\Omega_i$ is the driving strength with $\abs{\Omega_i}\to0$, and $\omega_d$ is the driving frequency. In physics, the driven process could be viewed as a coherent state input to the local system through a heat bath coupled to the corresponding local system \cite{SM2}. After tracing over the heat bath degrees of freedom, the evolution of the reduced density matrix $\rho_s$ is given by the Lindblad master equation \cite{Lme1,Lme2}
	\begin{align}\label{eq4}
		\frac{\dd\rho_s}{\dd t}=-i[H_{\text{sys}}+H_{\text{d}},\rho_s]+\sum_{k}(\kappa_{b,k}+\kappa_{c,k})\mathcal{D}[o_k]\rho_s,
	\end{align}
	where $\kappa_{b,k}=2\pi\abs{\xi_{b,k}}^2$, $\kappa_{c,k}=2\pi\abs{\xi_{c,k}}^2$, and $\mathcal{D}[a]\rho=a\rho a^\dagger-\{a^\dagger a,\rho\}/2$.
	
	Notice that Eq.~(\ref{eq4}) corresponds to the result of a zero-temperature limit (i.e., $\text{n}_\text{th}=0$) due to the fact that all heat baths are initially in the vacuum state, rather than thermal state. Now, according to the Mollow transformation \cite{PhysRevA.12.1919}, we can take off the driving term [i.e., Eq.~(\ref{eq3})], but the initial state of the heat bath coupled to the $i$-th local system must be replaced from the initial vacuum state to a coherent state. The heat bath mode can be $b_i$ or $c_i$. For convenience, we choose $b_i$ as the heat bath mode, and the corresponding heat bath is defined as an input channel. Similarly, the heat bath with mode $c_j$ is naturally defined as an output channel. Notice that the choice of $b_i$ and $c_j$ depends completely on the system we focus on and the physical quantities we calculate. As a consequence, the input state (i.e., the initial state of the total Hamiltonian) can be written as \cite{SM2}
	\begin{align}\label{eq5}
		|\psi_{\text{in}}\rangle=\mathscr{N}\sum_{n=0}^\infty\frac{\beta_i^n}{\sqrt{n!}}|\Psi^{(n)}_{\text{in}}\rangle^{b_i}_{\omega_d}\otimes|0\rangle_{\text{B}}\otimes|g\rangle,
	\end{align}
	where $|\Psi^{(n)}_{\text{in}}\rangle^{b_i}_{\omega_d}=b_i^{\dagger n}(\omega_d)/\sqrt{n!}|0\rangle$, which denotes the Fock state of $n$ photons with frequency $\omega_d$ in the input channel, $|0\rangle_\text{B}$ represents the vacuum state of baths except the mode $b_i$, and $|g\rangle$ is the ground state of the system, i.e., $H_{\text{sys}}|g\rangle=0$. Here, $\beta_i$ is the coherent state amplitude, i.e., $\beta_i=\Omega_i\sqrt{2\pi/\kappa_{b,i}}$, and $\mathscr{N}$ is the normalization factor. In this paper, the coherent state is abbreviated to $|\beta_i\rangle_{\omega_d}^{b_i}$, i.e., $|\psi_{\text{in}}\rangle=|\beta_i\rangle_{\omega_d}^{b_i}\otimes|0\rangle_{\text{B}}\otimes|g\rangle$.
	
	Based on Eqs.~(\ref{eq1}), and (\ref{eq5}) and scattering matrix ($S$-matrix) methods, we strictly prove an equation in Appendix \ref{B}, which relates the Lindblad master equation to the $S$ matrix through the $n$th-order equal-time correlation function (ETCF) of $j$-th local system 
	\begin{align}
		g_{jj}^{(n)}(0)=\frac{\langle\psi_{\text{out}}|c_j^{\dagger n}(t)  c_j^n(t)|\psi_{\text{out}}\rangle}{\ \ \langle\psi_{\text{out}}| c^\dagger_j(t)c_j(t)|\psi_{\text{out}}\rangle ^n}=
		\frac{\text{Tr}[o_j^{\dagger   n}o_j^n\rho_{ss}]}{\ \text{Tr}[o_j^{\dagger}o_j\rho_{ss}]^{n}} \label{eq6},
	\end{align}
	where $c_j(t)$ is the inverse Fourier transform of $c_j(\omega)$, i.e., $c_j(t)=\int d\omega/\sqrt{2\pi} e^{-i\omega t}c_j(\omega)\equiv\mathscr{F}^{-1}[c_j(\omega)]$, and $\rho_{ss}$ denotes the steady-state density matrix in Eq.~(\ref{eq4}). We construct a link between the input state and output state through the $S$ matrix, i.e., $|\psi_{\text{out}}\rangle=S|\psi_{\text{in}}\rangle$, where $S$ is the scattering operator \cite{SM9}. 
	
	Therefore, we connect Eq.~(\ref{eq4}) with Eq.~(\ref{eq1}) through the correlation function. That is to say, when we compute ETCF for some complex models with large dimensions of Hilbert spaces, we could replace the master-equation method with a more effective method, namely, $S$ matrix, and then take advantage of Eq.~(\ref{eq6}). Note that the method is not omnipotent. For example, it is not applicable in strong coherent driven situation and the case where the U(1) symmetry is not satisfied. On the contrary, it is highly reasonable to consider many models that satisfy the U(1) symmetry and involve a weakly coherent drive in quantum optics.
	\section{Scattering matrix and effective Hamiltonian}\label{III}
	In the above description, we have mentioned the scattering operator $S$, which is used to connect the input and output states, and the operator is equivalently written as $S=\Omega_-^\dagger\Omega_+$\,\cite{mwo1, mwo2}, where $\Omega_{\pm}=\exp({iH_{\text{tot}}t_{\pm}})\exp({-iH_{\text{B}}t_{\pm}})$ with $t_{\pm}\to\mp\infty$, called the Møller wave operators. Now, we consider $n$ photons scattering processes: $n$ photons incident from the input channel $\mu=(\mu_1,\mu_2,\ldots,\mu_n)$ with frequencies of $k=(k_1,k_2,\ldots,k_n)$ each, are scattered into the output channel $\nu=(\nu_1,\nu_2,\ldots,\nu_n)$ with frequencies of $p=(p_1,p_2,\ldots,p_n)$ each. According to the physical significance of the scattering operator, we define a generalized $n$-photon $S$ matrix with elements of the form:
	\begin{align}
		S^{\mu\nu}_{p_1\ldots p_n;k_1\ldots k_n}\ &= \ _\nu\langle p_1\cdots p_n|S|k_1\cdots k_n\rangle_\mu\nonumber\\
		\ &=\ \langle0|[\prod_{l=1}^{n}\nu_l(p_l)]S[\prod_{l=1}^{n}\mu_l^\dagger(k_l)]|0\rangle,\label{eq7}
	\end{align}
	where $|k_1\cdots k_n\rangle$ denotes the $n$-photon input state with frequencies $k$, $|p_1\cdots p_n\rangle$ denotes the $n$-photon outgoing state with frequencies $p$, and $\mu$ and $\nu$ represent the input and output channels, respectively. 
	
	Following the standard procedure \cite{PhysRevA.31.3761,SM9}, we obtain the input-output relations
	\begin{align}
		\mu_{i,\text{out}}(t)&=\mu_{i,\text{in}}(t)-io_{\mu_i}(t),\nonumber\\
		\nu_{j,\text{out}}(t)&=\nu_{j,\text{in}}(t)-io_{\nu_j}(t),\label{eq8}
	\end{align}
	where the input operator satisfies the commutation relation, i.e., $[\mu_{i,\text{in}}(t),\nu_{j,\text{in}}^\dagger(t^\prime)]=\delta_{\mu_i,\nu_j}\delta(t-t^\prime)$, and $o_{\mu_i}$ represents the annihilation operator of local system, which interacts with the channel $\mu_i$. Then, combining with Eq.~(\ref{eq7}) and Eq.~(\ref{eq8}), we have
	\begin{eqnarray}
		\begin{aligned}\label{eq9}
			S^{\mu\nu}_{p_1\ldots p_n;k_1\ldots k_n}= \langle0|[\prod_{l=1}^{n}\nu_{l,\text{out}}(p_l)][\prod_{l=1}^{n}\mu^\dagger_{l,\text{in}}(k_l)]|0\rangle,
		\end{aligned}
	\end{eqnarray}
	where $\mu_{l,\text{in}}(k)$ and $\nu_{l,\text{out}}(p)$ are the Fourier transformation of $\mu_{l,\text{in}}(t)$ and $\nu_{l,\text{out}}(t)$, respectively. The specific calculation details of Eqs.~(\ref{eq8}) AND (\ref{eq9}) are presented in Appendix \ref{A}.
	
	To calculate the $S$-matrix element analytically, we need to first derive the time-domain $S$-matrix element
	\begin{eqnarray}
		\begin{aligned}\label{eq10}
			S^{\mu\nu}_{t_1^\prime\ldots t_n^\prime;t_1,\ldots t_n}\equiv \langle0|[\prod_{l=1}^{n}\nu_{l,\text{out}}(t_l^\prime)][\prod_{l=1}^{n}\mu^\dagger_{l,\text{in}}(t_l)]|0\rangle,
		\end{aligned}
	\end{eqnarray}
	which is the inverse Fourier transform of Eq.~(\ref{eq9}). Subsequently, following Ref.\,\cite{SM1} and Eq.~(\ref{eq8}), the $n$-photon time-domain $S$-matrix element~(\ref{eq10}) is given by 
	\begin{widetext}
		\begin{align}
			S^{\mu\nu}_{t_1^\prime\ldots t_n^\prime;t_1\ldots t_n}=\sum_{m=0}^n\sum_{B_m,D_m}\sum_{P_c}G^{\mu_{D_m}\nu_{B_m}}(t^\prime_{B_m};t_{D_m}) \prod_{s=1}^{n-m}[\delta(t^\prime_{B^c_m(s)}-t_{P_cD^c_m(s)})\delta_{\nu_{B^c_m(s)},\mu_{P_cD^c_m(s)}}],\label{eq11}
		\end{align}
	\end{widetext}
	where the time-ordered $2n$-point Green's function is defined as
	\begin{align}
		G^{\mu\nu}(t^\prime_{B_n};t_{D_n})\equiv(-1)^n
		\langle0|\mathcal{T}[\prod_{l=1}^{n}o_{\nu_l}(t^\prime_l)o^\dagger_{\mu_l}(t_l)]|0\rangle,\label{eq12}
	\end{align}
	where $\mathcal{T}$ is the time-ordered symbol. In the expression above, $B_m$ [$D_m$] is a subset with $m$ elements of $\{1,\ldots,n\}$, its corresponding complementary subset is denoted by $B_m^c$ [$D_m^c$], $B_m(s)$ [$D_m(s)$] represent its $s$th element, and $P_cD_m^c$ is permutation over the subset $D_m^c$.  $\sum_{B_m}$ and $\sum_{P_c}$ represent a summation over all subsets with $m$ elements of $\{1,\ldots,n\}$ and all possible permutations $P_c$ of $D_m^c$, respectively. Besides, we use the shorthand notations $t^\prime_{B_m}\equiv\{t^\prime_i|i\in B_m\}$, $t_{D_m}\equiv\{t_i|i\in D_m\}$, $\nu_{B_m}\equiv(\nu_i|i\in B_m)$, and $\mu_{D_m}\equiv(\mu_i|i\in D_m)$. Due to $B_n=D_n=\{1,\ldots,n\}$, we have $\mu=\mu_{D_n}$ and $\nu=\nu_{B_n}$.
	
	For the case of different input and output channels, i.e., $\forall i,j\in\{1,2,\ldots,n\}$, $\mu_i\neq\nu_j$, the $n$-photon S-matrix is equal to the  $2n$-point Green's function due to the presence of coefficient $\delta_{\mu_i, \nu_j}$, i.e., $S^{\mu\nu}_{t^\prime_1\ldots t^\prime_n;t_1\ldots t_n}=G^{\mu\nu}(t^\prime_{B_n};t_{D_n})$. For the other case, i.e., $\exists i,j\in\{1,2,\ldots,n\}$, $\mu_i=\nu_j$, the $n$-photon S-matrix can be expressed as a sum over all possible products of a single Green's function and $\delta$ functions [see Eq.~(\ref{eq11})]. In effect, the $n$-photon S-matrix of the two cases can be explained by an intuitive physical process: if the incoming $n$ photons from one channel are scattered into another, all photons must first enter the local system before being scattered into another channel, and it means that this term containing “bare” $\delta$ functions in Eq.~(\ref{eq11}) will disappear. Conversely, if some of the input channels overlap with the output channels, some photons may bypass the local system but do not change themselves frequencies, i.e., $\delta(t^\prime_{B^c_m(s)}-t_{P_cD^c_m(s)})$, freely propagating from the input to output channel, i.e., $\nu_{B^c_m(s)}=\mu_{P_cD^c_m(s)}$, thereby naturally allowing the presence of “bare” $\delta$ functions.
	
	Finally, considering the infinite multiple degrees of freedom for heat baths, we have to seek an effective Hamiltonian which only contains the local system part ($H_{\text{sys}}\{o_k\}$) to replace the total Hamiltonian $H_{\text{tot}}$. In effect, the time-ordered Green's function is precisely computed by the effective Hamiltonian, not the whole system, and the derivation is presented in Appendix \ref{C}. The main result is
	\begin{align}
		G^{\mu\nu}(t^\prime_{B_n} ;t_{D_n})=\widetilde{G}^{\mu\nu}(t^\prime_{B_n} ;t_{D_n}),\label{eq13}
	\end{align}
	where 
	\begin{align}\label{eq14}
		\widetilde{G}^{\mu\nu}(t^\prime_{B_n} ;t_{D_n})=(-1)^n\langle g| \mathcal{T}[\prod_{l=1}^{n}\tilde{o}_{\nu_l}(t^\prime_l)\tilde{o}^\dagger_{\mu_l}(t_l)]|g\rangle,
	\end{align}
	with operators
	\begin{eqnarray}\label{eq15}
		\begin{bmatrix}
			\tilde{o}_{\nu_l}(t)\\
			\tilde{o}^\dagger_{\mu_l}(t)
		\end{bmatrix}=\exp(iH_{\text{eff}}t)
		\begin{bmatrix}
			o_{\nu_l}\\
			o^\dagger_{\mu_l}
		\end{bmatrix}
		\exp(-iH_{\text{eff}}t),
	\end{eqnarray}
	where
	\begin{eqnarray}
		\begin{aligned}\label{eq16}
			H_{\text{eff}}=H_{\text{sys}}\{o_k\}-\frac{i}{2}\sum_k(\kappa_{b,k}+\kappa_{c,k})o^\dagger_ko_k
		\end{aligned}
	\end{eqnarray}
	represents the effective Hamiltonian of the total system. We prove the effective Hamiltonian in Appendix \ref{C}. The effective Hamiltonian also commutes with the total excitation number operator, i.e., $[H_{\text{eff}}, \mathcal{N}]=0$. Hence, we convert the effective Hamiltonian to a block-diagonal form and the annihilation operator to block-upper triangular within the total excitation space: $H_{\text{eff}}\equiv{\rm{diag}}[\textbf{H}^{(0)}_{\text{eff}},\textbf{H}^{(1)}_{\text{eff}},\textbf{H}^{(2)}_{\text{eff}},\ldots]$ and
	\begin{equation}\label{eq17}
		o_{\mu_k}\!\equiv\! \begin{bmatrix}
			0&		\textbf{O}^{\mu_k}_{0,1}&		0&		\cdots\\
			0&0		&	\textbf{O}^{\mu_k}_{1,2}	&\cdots		\\
			0&	0	&	0	&	\cdots	\\
			\vdots&	\vdots	&\vdots		&		\ddots\\
		\end{bmatrix} \!,o^\dagger_{\mu_k}\!\equiv\! \begin{bmatrix}
			0&		0&		0&		\cdots\\
			\textbf{O}^{\dagger\mu_k}_{0,1}&0		&	0	&\cdots		\\
			0&	\textbf{O}^{\dagger\mu_k}_{1,2}	&	0	&	\cdots	\\
			\vdots&	\vdots	&\vdots		&		\ddots\\
		\end{bmatrix}.
	\end{equation}
	Here, the projection of the effective Hamiltonian on the $i$-th excitation subspace is denoted as $\textbf{H}_{\text{eff}}^{(i)}$, and the projection of the annihilation operator $o_{\mu_k}$ onto the direct sum of the $i$-th and $(i+1)$-th excitation subspace is denoted as $\textbf{O}^{\mu_k}_{i,i+1}$.
	\section{Probability amplitude of equal-time probing multiple photons}\label{IV}
	Inspired by the concepts of path integral in quantum mechanics, we define a probability amplitude of non-equal-time probing $n$ photons, i.e.,
	\begin{align}
		P_n^{\mu\nu}(t_1,\ldots,t_n)&\equiv\langle0|[\prod_{l=1}^{n}\nu_{l}(t_l)]S[\prod_{l=1}^{n}\mu^\dagger_{l}(k_l)]|0\rangle\nonumber\\
		&=\prod_{l=1}^{n}\int\frac{\dd{t^\prime_{l}}}{\sqrt{2\pi}}e^{-ik_{l}t^\prime_{l}}S^{\mu\nu}_{t_1\ldots t_n;t^\prime_1\ldots t^\prime_n}.\label{eq18}
	\end{align}
	Notably, the probability amplitude closely relates to Eq.~(\ref{eq7}), and it represents the probability amplitude of non-equal-time probing outgoing $n$ photons with frequencies $k$. Subsequently, in order to calculate the $n$th-order ETCF, we use a shorthand notation for equal-time probing case, i.e., $P_n^{\mu\nu}(t)\equiv P_n^{\mu\nu}(t,\ldots,t)$. We analytically derive the probability amplitude in Appendix \ref{D}, which is
	\begin{widetext}
		\begin{align}
			P^{\mu\nu}_n(t)=\frac{e^{-ik_{\text{tot}}t}}{\sqrt{(2\pi)^n}}\sum_{m=0}^n\sum_{D_m,B_m}\sum_{P,P_c}[\overrightarrow{\prod}_{j=1}^{m}\textbf{O}^{\nu_{B_m(j)}}_{j-1,j}][\overleftarrow{\prod}_{j={1}}^{m}\mathcal{K}_{\sum_{s=1}^{j}{k_{PD_m(s)}}}^{-1}(j)\textbf{O}^{\dagger \mu_{PD_m(j)}}_{j-1,j}]\prod_{s=1}^{n-m}[\delta_{\nu_{B^c_m(s)},\mu_{P_cD^c_m(s)}}],\label{eq19}
		\end{align}
	\end{widetext}
	where $\mathcal{K}_\omega(n)=-i[\textbf{H}_{\text{eff}}^{(n)}-\omega-i0^+]$, $\overleftarrow{\prod}_{j={1}}^{n}M_j=M_n\cdots M_2M_1$, $\overrightarrow{\prod}_{j={1}}^{n}M_j=M_1M_2\cdots M_n$, and $k_{\text{tot}}$ represents the total frequencies of $n$ incoming photons, i.e., $k_{\text{tot}}=\sum_{j=1}^nk_{j}$. The other symbols have been defined above, such as $B_m$, $D_m$, and $P_c$. Note that $PD_m$ is permutation over the subset $D_m$, and $\sum_{P}$ represents a summation over all possible permutations $P$ of $D_m$.
	
	\section{The classification based on input and output channels}\label{V}
	\begin{figure}
		\includegraphics[width=8.6cm]{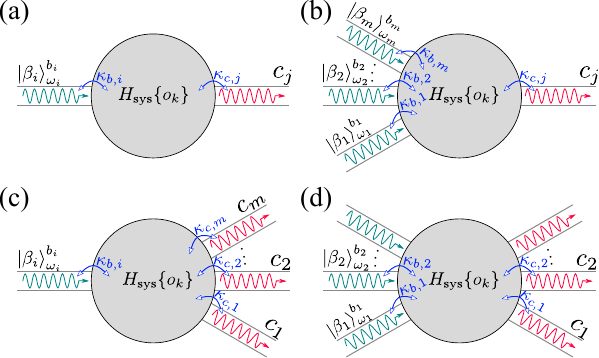}\\
		\caption{A concise schematic of input-output channels coupled to a system described by the Hamiltonian $H_{\text{sys}}\{o_k\}$. $\left({\rm{a}}\right)$ Single input channel and single output channel. $\left({\rm{b}}\right)$ Multiple input channels and single output channel. $\left({\rm{c}}\right)$ Single input channel and multiple output channels. $\left({\rm{d}}\right)$ Multiple input channels and multiple output channels. The cyan arrow represents the incoming coherent state within the input channel $b_i$, the red arrow represents the outgoing state within the output channel $c_j$, as indicated by the direction of the two arrows, and $\kappa_{b,i}$ and $\kappa_{c,j}$ are the corresponding decay rates.
		}\label{fig1}
	\end{figure}
	This section delves into the physics that arises from classifying based on input and output channels. The classification can be divided primarily into two categories. The first category consists of different input and output channels, which can be classified into four cases: one-to-one, many-to-one, one-to-many, and many-to-many relationships between the number of input and output channels, as shown in Figs.~\ref{fig1}(a)–\ref{fig1}(d). Here, we will concretely introduce the one-to-one, many-to-one, and one-to-many cases in Sec.~\ref{VA}, \ref{VB}, and \ref{VC}, respectively. Similarly, the many-to-many case could be simply obtained by combining many-to-one and one-to-many cases. 
	
	The second category consists of the same input and output channels and applies mainly to waveguide quantum electrodynamics (QED)\,\cite{Shen:05, wg7,wg6,wg5,wg4,wg3,wg1} systems. This emerging field focuses on the interaction of propagating photons with quantum dots \cite{qd3,qd1,qd2}, superconducting qubits \cite{qb1,qb2,qb5,qb3,qb4}, or nanocavities \cite{ncavity1,ncavity2,ncavity3,ncavity4}. We will discuss this category in detail in Sec.~\ref{VD}.
	
	\subsection{Single input and single output channels}\label{VA}
	In Sec.~\ref{II}, we have introduced the representative one-to-one case. Thus, for the incoming $n$ photons, we have $\mu=(b_i)^n$, $\nu=(c_j)^n$, and $k=(\omega_d)^n$, where the superscript $n$ represents repeating number of each element in the list, e.g, $(b_i)^2=(b_i,b_i)$. The corresponding input-output formalism can be written as
	\begin{align}
		b_{i,\text{out}}(t)&=b_{i,\text{in}}(t)-io_{b_i}(t),\label{eq20}\\
		c_{j,\text{out}}(t)&=c_{j,\text{in}}(t)-io_{c_j}(t)\label{eq21},
	\end{align}
	where $o_{b_i}(t)=\sqrt{\kappa_{b,i}}o_i(t)$ and $o_{c_j}(t)=\sqrt{\kappa_{c,j}}o_j(t)$. Notably, the only non-zero term is for $m=n$ in Eq.~(\ref{eq19}); therefore, the probability amplitude of equal-time probing $n$-photon could be further simplified as 
	\begin{align}
		P_n^{\mu\nu}(t)=n!\xi^n[\overrightarrow{\prod}_{l=1}^{n}\textbf{O}^j_{l-1,l}][\overleftarrow{\prod}_{l=1}^{n}\mathcal{K}_{l\omega_d}^{-1}(l)\textbf{O}^{\dagger i}_{l-1,l}],\label{eq22}
	\end{align}
	where $\xi=\sqrt{\kappa_{b,i}\kappa_{c,j}/2\pi}\exp({-i\omega_d t})$.
	
	Then, the $n$th-order ETCF~(\ref{eq6}) under a weak coherent drive is given by
	\begin{align}
		\tilde{g}_{jj}^{(n)}(0)&\equiv\lim\limits_{\abs{\beta_i}\to0}g_{jj}^{(n)}(0)=\frac{|P_{n}^{\mu\nu}(t)/n!|^2}{\ \ |P_1^{\mu\nu}(t)|^{2n}}\nonumber\\
		&=\frac{|[\overrightarrow{\prod}_{l=1}^{n}\textbf{O}^j_{l-1,l}][\overleftarrow{\prod}_{l=1}^{n}\mathcal{K}_{l\omega_d}^{-1}(l)\textbf{O}^{\dagger i}_{l-1,l}]|^2}{|\textbf{O}_{0,1}^{j}\mathcal{K}_{\omega_d}^{-1}(1)\textbf{O}_{0,1}^{\dagger i}|^{2n}}.\label{eq23}
	\end{align}
	Besides, an essential physical quantity, namely single-photon transmission is
	\begin{align}
		T&\equiv\lim\limits_{\abs{\beta_i}\to0}\frac{\langle\psi_{\text{out}}|c^{\dagger }_j(t) c_j(t)|\psi_{\text{out}}\rangle}{\langle\psi_{\text{in}}| b_i^\dagger(t)b_i(t)|\psi_{\text{in}}\rangle }=2\pi\abs{P_1^{\mu\nu}}^2\nonumber\\
		&= \kappa_{b,i}\kappa_{c,j}|\textbf{O}_{0,1}^{j}\mathcal{K}_{\omega_d}^{-1}(1)\textbf{O}_{0,1}^{\dagger i}|^{2}.\label{eq24}
	\end{align}
	The detailed derivation procedure is provided at the end of Appendix \ref{D}.
	
	\subsection{Multiple input and single output channels}\label{VB}
	According to Eq.~(\ref{eq7}) structure, the $S$-matrix elements are formally the same to the two categories, but the difference is their superscripts: $\mu$ and $\nu$. Since the possible scattering paths of multiple photons present multiple choices when multiple input and output channels are present, so each scattering path corresponds to a unique $\mu$ and $\nu$ in terms of the $S$-matrix. For example, in the one-to-one case, $n$ photons can only be injected through a single input channel $b_i$ and scattered into a single output channel $c_j$, so the possible scattering path of $n$ photons is only one, i.e., $\mu=(b_i)^n$ and $\nu=(c_j)^n$, where the number of elements both are $n$. Meanwhile, the frequency of each incoming photon is naturally determined when $\mu$ is determined, i.e., incoming $n$ photons frequencies $k=(k_1,k_2,\ldots,k_n)$.
	
	As shown in Fig.~\ref{fig1}(b), there are multiple input channels where each channel is injected with a coherent state, and the input state can be written as
	\begin{align}\label{eq25}
		|\psi_{\text{in}}\rangle=[\prod_{l=1}^{m}|\beta_l\rangle_{\omega_l}^{b_l}]|0\rangle_{\text{B}}|g\rangle,
	\end{align}
	where $|0\rangle_{\text{B}}$ represents the vacuum state of baths except for multiple input channels, and $m$ is the number of channels injected coherent state. According to the Mollow transformation, the coherent state input is equivalent to the Hamiltonian, namely a multi-mode coherent drive,
	\begin{align}\label{eq26}
		H_{\text{d}}=\sum_{l=1}^m\frac{\beta^*_l o_le^{i\omega_{l}t}+\beta_l o^\dagger_le^{-i\omega_{l}t}}{\sqrt{2\pi/\kappa_{b,l}}}.
	\end{align}
	
	On the one hand, we note that the Hamiltonian, $H_{\text{sys}}\{o_k\}+H_{\text{d}}$, is always time-dependent when these driving frequencies are not all identical (i.e., $ \exists k,l,\omega_k\neq\omega_l$), and its reduced density matrix also is time-dependent even after long-time evolution, i.e., $\rho_s(t_0+t)|_{t_0\to+\infty}\neq\rho_{ss}$. On the other hand, the Hamiltonian can be transformed into a time-independent one by selecting an applicable rotating frame when all driving frequencies are identical, and we have $\rho_s(t_0+t)|_{t_0\to+\infty}=\rho_{ss}$. As a result, Eq.~(\ref{eq6}) needs to corrected slightly to apply simultaneously to both cases, i.e.,
	\begin{align}
		g_{jj}^{(n)}(t)&=\frac{\langle\psi_{\text{out}}|c_j^{\dagger n}(t)  c_j^n(t)|\psi_{\text{out}}\rangle}{\ \ \langle\psi_{\text{out}}| c^\dagger_j(t)c_j(t)|\psi_{\text{out}}\rangle ^n}\nonumber\\
		&=\frac{\text{Tr}[o_j^{\dagger   n}o_j^n\rho_s(t_0+t)]}{\ \ \text{Tr}[o_j^{\dagger}o_j\rho_s(t_0+t)]^{n}}\bigg|_{t_0\to+\infty} \label{eq27}.
	\end{align}
	
	From now on, we assume $\beta_l=\eta_l\beta_1$ for all $l$. According to the discussions above, the computation of $n$th-order ETCF under the weak coherent state amplitude ($\abs{\beta_1}\to0$) is as follows:
	\begin{align}
		&\tilde{g}_{jj}^{(n)}(t)\equiv\lim\limits_{\abs{\beta_1}\to0}g_{jj}^{(n)}(t)=\frac{|\sum^{(n,m)}_{\mu}P_{n}^{\mu\nu}(t)c_\mu|^2}{\ \ |\sum^{(1,m)}_{\mu}P_1^{\mu\nu}(t)c_\mu|^{2n}}\nonumber\\
		&=\frac{|[\overrightarrow{\prod}_{l=1}^{n}\textbf{O}^{j}_{l-1,l}][\sum_{x_1,\ldots,x_n=1}^{m}\overleftarrow{\prod}_{l=1}^{n}\mathcal{K}_{\sum_{i=1}^l\omega_{x_i}}^{-1}(l)\overline{\textbf{O}}^{\dagger }_{l-1,l}]|^2}{|\textbf{O}_{0,1}^{j}\sum_{x_1=1}^m\mathcal{K}_{\omega_{x_1}}^{-1}(1)\overline{\textbf{O}}_{0,1}^{\dagger }|^{2n}}.\label{eq28}
	\end{align}
	
	In the first step above, we redefine Eq.~(\ref{eq27}) under the $m$ weak coherent state amplitudes.
	In the second step, we first take advantage of Eqs.~(\ref{eq18}), (\ref{eq25}), and (\ref{eq27}). Then, we define the summation $\sum^{(n,m)}_{\mu}$ and combination coefficient $c_\mu$. Here, $\sum^{(n,m)}_{\mu}$ represents a summation over all sublists with $n$ elements of $(b_1,b_2,\ldots,b_m)^n$, and $c_\mu=\prod_{k=1}^{m}[\eta_k^{z_\mu(b_k)}/z_\mu(b_k)!]$, where $z_\mu(b_k)$ represents the number of $b_k$ in list $\mu$. For example, when $n=2$ and $m=2$, there are three corresponding sublists $\mu_1=(b_1)^2,\mu_2=(b_2)^2,\mu_3=(b_1,b_2)$, and $z_{\mu_1}(b_1)=z_{\mu_2}(b_2)=2, z_{\mu_1}(b_2)=z_{\mu_2}(b_1)=0,z_{\mu_3}(b_1)=z_{\mu_3}(b_2)=1$.
	In the last step, we use the probability amplitude expression~(\ref{eq19}), and define the time-dependent symbol, i.e.,  $\overline{\textbf{O}}_{l-1,l}^\dagger=\sum_{j=1}^m[\eta_j\textbf{O}_{l-1,l}^{\dagger b_j}e^{-i\omega_j t}\delta_{j,x_l}]$.
	
	We find that the result of Eq.~(\ref{eq28}) applies to the two cases above. Meanwhile, for the case of $\omega_j=\omega_d$, for all $j$, i.e., the frequencies of all incoming photons are equal to $\omega_d$, the $n$th-order ETCF could be further simplified as
	\begin{align}
		\tilde{g}_{jj}^{(n)}(t)
		&=\frac{|[\overrightarrow{\prod}_{l=1}^{n}\textbf{O}^{j}_{l-1,l}][\overleftarrow{\prod}_{l=1}^{n}\mathcal{K}_{l\omega_d}^{-1}(l)\sum_{i=1}^m{\eta_i\textbf{O}_{l-1,l}^{\dagger b_i}}]|^2}{|\textbf{O}_{0,1}^{j}\mathcal{K}_{\omega_d}^{-1}(1)\sum_{i=1}^m{\eta_i\textbf{O}_{0,1}^{\dagger b_i}}|^{2n}}\nonumber\\
		&=\tilde{g}_{jj}^{(n)}(0).\label{eq29}
	\end{align}
	Similarly, the single-photon transmission is given by
	\begin{align}
		T&=\lim\limits_{\abs{\beta_1}\to0}\frac{\langle\psi_{\text{out}}|c^{\dagger }_j(t) c_j(t)|\psi_{\text{out}}\rangle}{\sum_{i=1}^m\langle\psi_{\text{in}}|b_i^\dagger(t)b_i(t)|\psi_{\text{in}}\rangle}\nonumber\\
		&
		= \kappa_{c,j}\frac{|\textbf{O}_{0,1}^{j}\mathcal{K}_{\omega_d}^{-1}(1)\sum_{i=1}^m\eta_i\textbf{O}^{\dagger b_i}_{0,1}|^{2}}{\sum_{i=1}^m\abs{\eta_i}^2}.\label{eq30}
	\end{align}
	Note that $\eta_1=1$. As expected, when $\eta_i=0$, for $i>1$, the multi-mode coherent drive becomes back to the single-mode drive, and Eqs.~(\ref{eq29}) and (\ref{eq30}) also are in agreement with Eqs.~(\ref{eq23}) and (\ref{eq24}). 
	
	In conclusion, we study multiple input channels because they provide significant advantages in some physical problems. More specifically, when studying a conventional photon blockade model \cite{CPB2,CPB1,CPB4,CPB3}, such as the widely studied nonlinear system of a cavity filled with Kerr material \cite{PhysRevA.49.R20,PhysRevX.10.021022}, achieving a near-perfect photon blockade effect is crucial for quantum information processing and quantum computing. However, this needs a strong nonlinearity without other improved approaches. Based on the theory of multiple input channels, the nonlinearity strength required significantly decreases \cite{dyna_pb2} for the same strong photon blockade. The reason is that multiple input channels lead to multiple possible scattering processes for the outgoing two photons, and the probability amplitudes of these processes can interfere destructively. Consequently, the conventional photon blockade model enters the unconventional photon blockade \cite{UPB2,UPB1,UPB4,UPB3} region through interference between multiple input channels. Note in this model that we need to use the dynamical correlation function~(\ref{eq28}) to study dynamical photon blockade effect, which corresponds to the case, $\exists k,l,\omega_k\neq\omega_l$.
	\subsection{Single input and multiple output channels}\label{VC}
	As shown in Fig.~\ref{fig1}(c), considering the presence of multiple output channels, now we will study the photonic correlation between different output channels, and we could use the cross-correlation function \cite{cross-corr1,cross-corr2,cross-corr3} (CCF) with zero-delay time to characterize it, i.e.,
	\begin{align}
		g_{\nu}^{(n)}(0)&\equiv\frac{\langle\psi_{\text{out}}|[\prod_{j=1}^{n}\nu_j^{\dagger }(t) ][\prod_{j=1}^{n} \nu_j(t)]|\psi_{\text{out}}\rangle}{\ \ \prod_{j=1}^{n}\langle\psi_{\text{out}}| \nu^\dagger_j(t)\nu_j(t)|\psi_{\text{out}}\rangle}\nonumber\\
		&=\frac{\text{Tr}[(\prod_{j=1}^{n}o_{\nu_j}^{\dagger})(\prod_{j=1}^{n}o_{\nu_j})\rho_{ss}]}{\ \ \prod_{j=1}^{n}\text{Tr}[o_{\nu_j}^{\dagger}o_{\nu_j}\rho_{ss}]}.\label{eq31}
	\end{align}
	
	We find that the form of CCF agrees with the $n$th-order ETCF, so we can repeat the same as steps, like Sec.~\ref{VA}, to derive its analytical expression. Meanwhile, the probability amplitude~(\ref{eq18}) is valid for arbitrary $\mu$ and $\nu$. Thereby, the $n$-photon CCF under a weak coherent drive with driving frequency $\omega_d$ is given by
	\begin{align}
		\tilde{g}_{\nu}^{(n)}(0)&\equiv\lim\limits_{\abs{\beta_1}\to0}g_{\nu}^{(n)}(0)=\frac{|P_{n}^{\mu\nu}(t)/n!|^2}{\prod_{j=1}^{n}|P_1^{\mu\nu_j}(t)|^{2}}\nonumber\\
		&=\frac{|[\overrightarrow{\prod}_{l=1}^{n}\textbf{O}^{\nu_{l}}_{l-1,l}][\overleftarrow{\prod}_{l=1}^{n}\mathcal{K}_{l\omega_d}^{-1}(l)\textbf{O}^{\dagger i}_{l-1,l}]|^2}{\prod_{j=1}^{n}|\textbf{O}_{0,1}^{\nu_j}\mathcal{K}_{\omega_d}^{-1}(1)\textbf{O}_{0,1}^{\dagger i}|^{2}}.\label{eq32}
	\end{align}
	As expected, when the number of output channels is equal to one, i.e., $m=1$, Eq.~(\ref{eq32}) will become back to Eq.~(\ref{eq23}) due to $\nu=(c_j)^n$. 
	
	Analogously, for the case of multiple input channels and multiple output channels, as shown in Fig.~\ref{fig1}(d), we could directly derive the correlation function according to the two conclusions of Sec.~\ref{VB} with Sec.~\ref{VC}.
	\subsection{Same input and output channel}\label{VD}
	\begin{figure}[h]
		\includegraphics[width=8.6cm]{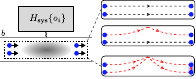}\\
		\caption{Illustration of two-photon scattering processes for a system side-coupled to a waveguide. There are three possible scattering paths for the two photons, corresponding to the descriptions in the three figures on the right. Here, we assume the input-output formalism is  $b_{\text{out}}(t)=b_{\text{in}}(t)-io_b(t)$.
		}\label{fig2}
	\end{figure}
	Based on the structure of $n$-photon $S$ matrix~(\ref{eq11}), we only consider the case of identical single input channel and single output channel here, i.e., $\mu=\nu=(b)^n$, and the rest of cases could be obtained by combining our proved conclusions. Here, we assume the frequency of the incoming coherent state is $\omega_d$. Thus, the probability amplitude of equal-time probing $n$-photon could be further simplified as 
	\begin{align}
		P_n^{\mu\mu}(t)=n!\mathcal{I}^{n}\sum_{m=0}^nC_n^m[\overrightarrow{\prod}_{l=1}^{m}\textbf{O}^{b}_{l-1,l}][\overleftarrow{\prod}_{l=1}^{m}\mathcal{K}_{l\omega_d}^{-1}(l)\textbf{O}^{\dagger b}_{l-1,l}],\label{eq33}
	\end{align}
	where $\mathcal{I}=\exp({-i\omega_d t})/\sqrt{2\pi}$, and $C_n^m$ is the combination number formula.
	
	Notably, $P_{n}^{\mu\mu}$ can be described by a natural and intuitive physical process, i.e., the $m$ incoming photons are scattered into the output channel $b$ by the local system, and the rest of $(n-m)$ photons bypass the local system, where $0\le m\le n$. The property is radically different from the case of multiple input channels above. From the perspective of interference paths, one is that $n$ photons have multiple possible scattering paths due to the presence of multiple input channels, and the other is that $n$ photons could be divided into multiple combinations between few-photon scattering paths and freely propagating paths due to the presence of identical input and output channel, as shown in Fig.~\ref{fig2}. 
	
	Similarly, the input state is $|\psi_{\text{in}}\rangle=|\beta\rangle_{\omega_d}^{b}|0\rangle_{B}|g\rangle$. Therefore, the $n$th-order ETCF and single-photon transmission under the weak coherent state amplitude are given by
	\begin{align}
		\tilde{g}^{(n)}_{bb}(0)&\equiv\lim\limits_{\abs{\beta}\to0}\frac{\langle\psi_{\text{out}}|b^{\dagger n}(t)  b^n(t)|\psi_{\text{out}}\rangle}{\ \ \langle\psi_{\text{out}}| b^\dagger(t)b(t)|\psi_{\text{out}}\rangle ^n}=\frac{\abs{P^{\mu\mu}_n(t){/n!}}^2}{\ \ \abs{P^{\mu\mu}_1(t)}^{2n}},\nonumber\\
		T\equiv&\lim\limits_{\abs{\beta}\to0}\frac{\langle\psi_{\text{out}}|b^{\dagger }(t)  b(t)|\psi_{\text{out}}\rangle}{\langle\psi_{\text{in}}| b^\dagger(t)b(t)|\psi_{\text{in}}\rangle }=2\pi\abs{P^{\mu\mu}_1(t)}^2.\label{eq34}
	\end{align}
	However, the form of effective Hamiltonian usually differs from Eq.~(\ref{eq16}) in waveguide QED systems because $o$ could consist of the annihilation operators of multiple local systems, e.g., $o_b=\xi_1o_1+\xi_2o_2$. The specific derivation processes refer to Refs.\,\cite{PhysRevA.91.042116,PhysRevA.95.063809,PhysRevResearch.2.013369}. Note here that we should do a standard quantum-optical Born-Markov approximation and neglect retardation effects.
	
	In general, the imaginary part of all eigenvalues of the effective Hamiltonian is less than or equal to zero for a pure dissipative system satisfying the U(1) symmetry, and the number of eigenvalues whose imaginary parts are equal to zero represents the number of steady states \cite{PhysRevA.89.052133}, such as the effective Hamiltonian~(\ref{eq16}), which its steady state is unique. However, multiple steady states are easy to achieve in a waveguide QED system, such as multiple local quantum systems coupled with a waveguide in different locations \cite{Albrecht_2019,PhysRevLett.122.203605,PhysRevX.10.031011}. Notably, our method is still effective for the case of multiple steady states since $\mathcal{K}_{\omega}(n)$ always is reversible due to the presence of $i0^+$.
	\begin{figure*}
		\includegraphics[width=17.8cm]{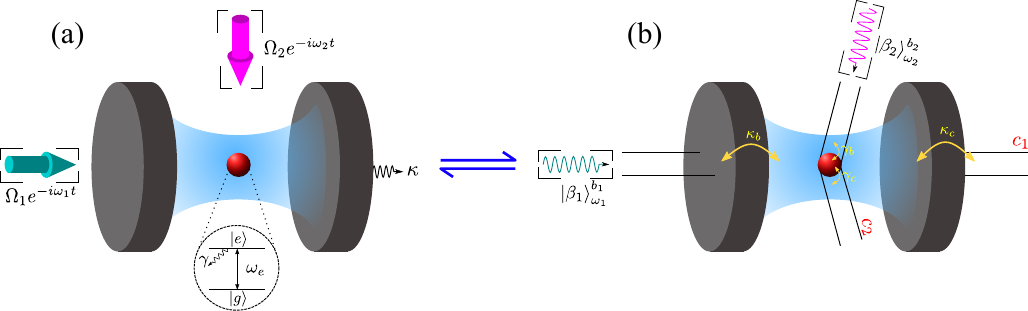}\\
		\caption{$\left({\rm{a}}\right)$ Sketch of a two-level atom with transition frequency $\omega_e$ trapped in a single-mode cavity. The cyan (magenta) arrow corresponds to the cavity (atom) drive with the driving strength $\Omega_1$ ($\Omega_2$) and frequency $\omega_1$ ($\omega_2$). $|g\rangle$ ($|e\rangle$) is the ground (excited) state of the atom, and $\kappa$ and $\gamma$ are the cavity and atomic decay rates, respectively. $\left({\rm{b}}\right)$ The equivalent model of $\left({\rm{a}}\right)$. Here, there are two input channels ($b_1$ and $b_2$) and output channels ($c_1$ and $c_2$). $\kappa_b$ ($\gamma_b$) and $\kappa_c$ ($\gamma_c$) are the decay rates of the cavity (atom) coupled to $b_1$ ($b_2$) and $c_1$ ($c_2$), respectively. $|\beta_1\rangle_{\omega_1}^{b_1}$ and $|\beta_2\rangle_{\omega_2}^{b_2}$ are the incoming coherent states, which correspond to the cavity-driven and atom-driven cases, respectively. Meanwhile, we have $\kappa=\kappa_b+\kappa_c$ ,$\gamma=\gamma_b+\gamma_c$, $\beta_1\sqrt{\kappa_b/2\pi}=\Omega_1$ and $\beta_2\sqrt{\gamma_b/2\pi}=\Omega_2$.
		}\label{fig3}
	\end{figure*}
	\section{Paradigmatic Examples}\label{VI}
	In this section, we will apply the method developed in the previous sections to the three typical examples. In Sec.~\ref{VIA}, we consider a typical single-atom–cavity QED system and then analyze three schemes: cavity-driven, atom-driven, and cavity-atom-driven. Besides, we also study the advantages of the multi-mode drive for the photon blockade. In Sec.~\ref{VIB}, we consider a coupled-cavity array system with multiple atoms. Undoubtedly, we will face dimensional exponential growth of the Hilbert space with the size of the cavity, so we have to resort to more effective numerical methods, such as the Monte Carlo method \cite{Qt}, tensor networks \cite{TN4,TN3,TN2,TN1,TN5}, or a combination of both \cite{manzoni_simulating_2017}. However, our approach could reduce this exponential complexity to polynomial complexity. In Sec.~\ref{VIC}, we consider a spin-$\frac{1}{2}$ system coupled to a one-dimensional (1D) waveguide model, thereby studying the relationship between the transmission spectrum and the correlation function.
	
	Finally, based on the compact analytical expressions mentioned above, we present a Python code that can handle any quantum systems satisfying certain conditions required in the paper. The Python notebooks containing the code for each physical model studied here are available online \cite{Qcs}.
	\subsection{Jaynes-Cummings Model}\label{VIA}
	In this section, we consider the most typical model in quantum optics: Jaynes-Cummings (JC) model \cite{JC}, which describes a two-level atom coupled to a single-mode cavity field. The Hamiltonian reads as
	\begin{align}\label{eq35}
		H_{\text{JC}}=\omega_e\sigma^\dagger \sigma + \omega_c a^\dagger a+\textsl{g}(\sigma^\dagger a+\sigma a^\dagger),
	\end{align}
	where $a$ and $\omega_c$ represent the annihilation operator and resonant frequency of the cavity mode, respectively; $\sigma$, $\textsl{g}$, and $\omega_e$ are the lowering operator ($|g\rangle\langle e|$), cavity coupling strength, and transition frequency of the atom. Then, we consider three schemes: cavity-driven ($H_{\text{d}}^c$), atom-driven ($H_{\text{d}}^e$), and cavity-atom-driven ($H_{\text{d}}^{ce}$). Using the notation $\mathcal{D}[A]\rho\equiv A\rho A^\dagger-\{A^\dagger A,\rho\}/2$, the master equation in explicit Lindblad form reads as ($\hbar=1$)
	\begin{align}\label{eq36}
		\dot{\rho}=-i[H_{\text{JC}}+H_{\text{d}}^i,\rho]+\kappa\mathcal{D}[a]\rho+\gamma\mathcal{D}[\sigma_-]\rho,
	\end{align}
	where $\kappa$ and $\gamma$ are the cavity and atomic decay rates, respectively, and $H_{\text{d}}^i$ is the driving term, e.g., $H_{\text{d}}^{c}=(\Omega_1^*ae^{i\omega_1t}+\text{H.c.})$, $H_{\text{d}}^{e}= (\Omega_2^*\sigma e^{i\omega_2t}+\text{H.c.})$, and $H_{\text{d}}^{ce}=H_{\text{d}}^c+H_{\text{d}}^e$. Now, we assume $\Omega_2=\eta\Omega_1$ and $\abs{\Omega_1}\to0$. On the one hand, we could obtain some useful information about the system by an equivalent Hamiltonian like Eq.~(\ref{eq1}), such as statistical properties of output light. On the other hand, the approach also provides a more intuitive physical process to multi-mode drive, as shown in Fig.~\ref{fig3}.
	
	Subsequently, following the standard procedures of our method, the effective Hamiltonian and the total excitation number operator are given by
	\begin{align}\label{eq37}
		H_{\text{eff}}=H_{\text{JC}}-\frac{i\kappa}{2}a^\dagger a-\frac{i\gamma}{2}\sigma^\dagger\sigma,\quad \mathcal{N}=a^\dagger a+\sigma^\dagger\sigma.
	\end{align}
	Thereby, by selecting a set of basis vectors, we have 
	\begin{align}
		\textbf{H}_{{\rm{eff}}}^{(n)} &= \begin{bmatrix}
			{n{{\tilde \omega }_{\rm{c}}}}&{{\rm{g}}\sqrt n }\\
			{{\rm{g}}\sqrt n }&{(n - 1){{\tilde \omega }_{\rm{c}}} + {{\tilde \omega }_{\rm{e}}}}
		\end{bmatrix}, \nonumber\\
		\;\textbf{O}_{n - 1,n}^e &= \begin{bmatrix}
			0&1\\
			0&0
		\end{bmatrix},\textbf{O}_{0,1}^e = \begin{bmatrix}
			0&1
		\end{bmatrix}, \label{eq38}\\
		\textbf{O}_{n - 1,n}^{\rm{c}} &= {\rm{diag}}\left[\sqrt{n} ,\sqrt{n-1}\right]  ,\;\textbf{O}_{0,1}^c = \begin{bmatrix}
			1&0
		\end{bmatrix},\;\nonumber
	\end{align}
	where $\textbf{O}^c_{n-1,n}$ ($\textbf{O}^e_{n-1,n}$) represent the projection of the annihilation operator $a$ ($\sigma$) onto the direct sum of the $(n-1)$-th and $n$-th excitation subspace, $\tilde{\omega}_c=\omega_c-i\kappa/2$, and $\tilde{\omega}_e=\omega_e-i\gamma/2$. 
	
		\begin{figure}
		\includegraphics[width=8.6cm]{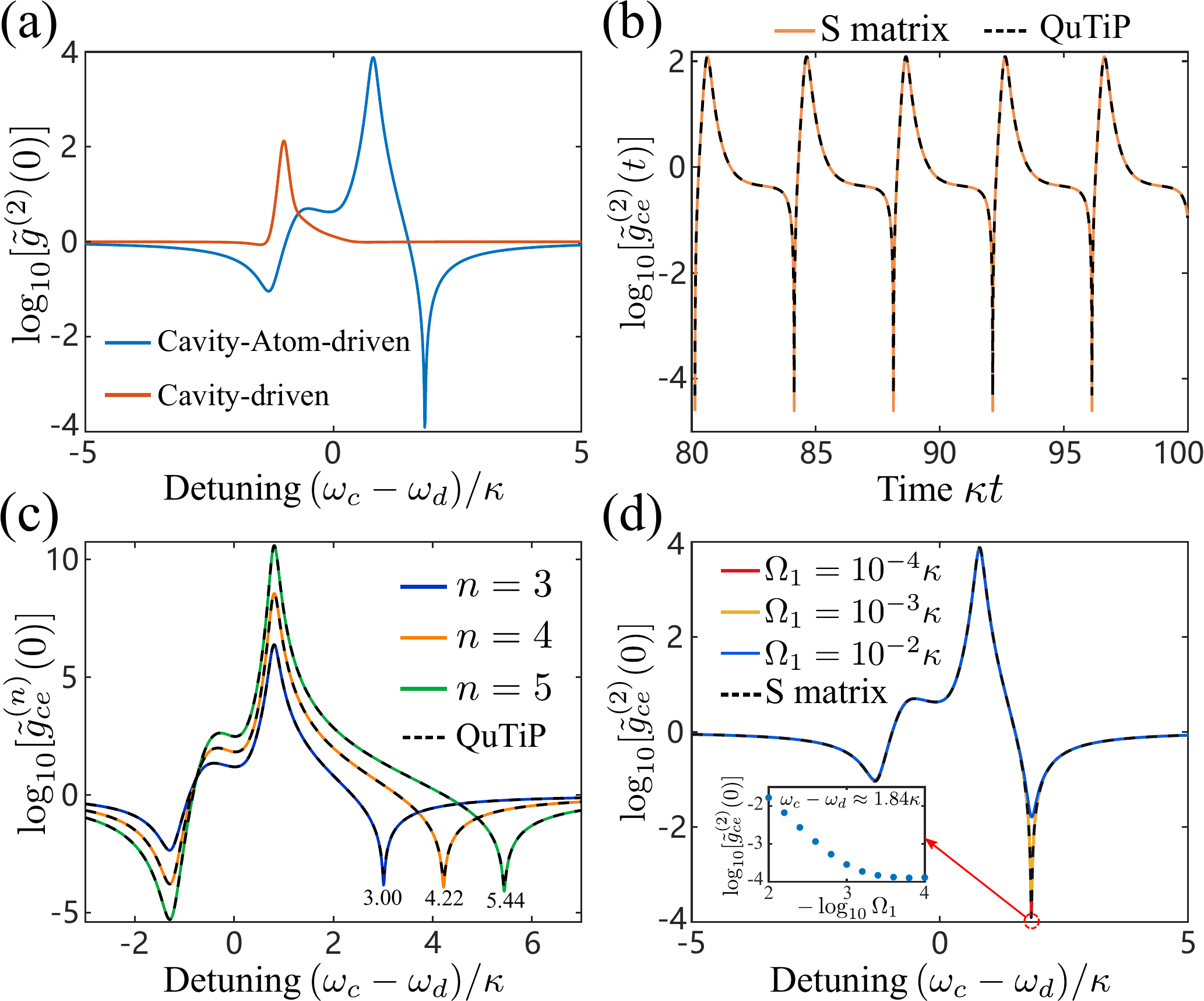}\\
		\caption{$\left({\rm{a}}\right)$ Equal-time second-order correlation $\log_{10}[\tilde{g}^{(2)}(0)]$ versus the detuning $(\omega_c-\omega_d)/\kappa$ for the two schemes: cavity-atom driven and cavity driven. $\left({\rm{b}}\right)$ Equal-time second-order dynamical correlation $\log_{10}[\tilde{g}_{ce}^{(2)}(t)]$ versus the time $\kappa t$. $\left({\rm{c}}\right)$ Equal-time $n$th-order correlation $\log_{10}[\tilde{g}^{(n)}_{ce}(0)]$ versus the detuning $(\omega_c-\omega_d)/\kappa$, where $n=3,4,5$. $\left({\rm{d}}\right)$ Validation of the $S$-matrix calculation against QUTIP for the computation of the second-order ETCF. The inset shows the dependence of $\log_{10}[\tilde{g}^{(2)}_{ce}(0)]$ as a function of the driving strength $\Omega_1$ in $\omega_c-\omega_d\approx1.84\kappa$. In all subplots, the system parameters are given by $\gamma=0.2\kappa$ and $\textsl{g}=0.6\kappa$. The other parameters for (a), (c), and (d) are chosen as $\omega_1=\omega_2=\omega_d$, $\omega_e-\omega_c=\kappa$, and $\Omega_2=3\Omega_1$; $\omega_c-\omega_1=-(\omega_c-\omega_2)=-\pi\kappa/4$, $\omega_c=\omega_e$, and $\Omega_2=\Omega_1$ for $\left({\rm{b}}\right)$. 
		}\label{fig4}
	\end{figure}
	
	When $\omega_1=\omega_2=\omega_d$, by plugging Eq.~(\ref{eq38}) into Eqs.~(\ref{eq23}) and (\ref{eq29}), we can obtain the analytical solution of $n$th-order ETCF about cavity mode. For the sake of conciseness, taking $n=2$ as an example, we have
	\begin{align}
		\tilde{g}_e^{(2)}(0)
		&=\frac{|\Delta_c\Delta_e-\textsl{g}^2|^2}{|\Delta_c(\Delta_c+\Delta_e)-\textsl{g}^2|^2},\nonumber\\
		\tilde{g}_c^{(2)}(0)
		&=g_e^{(2)}(0)\frac{|\Delta_e(\Delta_c+\Delta_e)+\textsl{g}^2|^2}{\abs{\Delta_e}^4}, \label{eq39}\\
		\tilde{g}_{ce}^{(2)}(0)
		&=g_e^{(2)}(0)\frac{|\textsl{g}^2-\Delta_c(\Delta_c+\Delta_e)+(\Delta_c+\Delta_e-\eta\textsl{g})^2 |^2}{|\Delta_e-\eta\textsl{g}|^4},\nonumber
	\end{align}
	where the subscripts $c$, $e$, and $ce$ represent the scheme of cavity driven, atom driven, and both, respectively, and $\Delta_{xj}=\tilde{\omega}_{x}-\omega_j=\Delta_{x}$. Besides, we also find the relationship by analyzing $\eta$, i.e., $\tilde{g}_{ce}^{(2)}(0)|_{\eta\to0}=\tilde{g}_c^{(2)}(0), \tilde{g}_{ce}^{(2)}(0)|_{\eta\to\infty}=\tilde{g}_e^{(2)}(0)$.
	
	In addition, when $\omega_1\neq\omega_2$, by plugging Eq.~(\ref{eq38}) into Eq.~(\ref{eq28}), the second-order dynamical ETCF is given by ($\delta=\omega_2-\omega_1$)
	\begin{align}\label{eq40}
		\tilde{g}_{ce}^{(2)}(t)=\frac{\abs{C_1+C_2e^{-2i\delta t}+C_3e^{-i\delta t}+C_4e^{-i\delta t}}^2}{\abs{C_5+C_6e^{-i\delta t}}^4},
	\end{align}
	with coefficients 
	\begin{align}
		C_1&=\frac{\Delta_{e1}(\Delta_{e1}+\Delta_{c1})+\textsl{g}^2}{[\Delta_{e1}\Delta_{c1}-\textsl{g}^2][\Delta_{c1}(\Delta_{e1}+\Delta_{c1})-\textsl{g}^2]},\nonumber\\
		C_2&=\frac{\eta^2\textsl{g}^2}{[\Delta_{e2}\Delta_{c2}-\textsl{g}^2][\Delta_{c2}(\Delta_{e2}+\Delta_{c2})-\textsl{g}^2]},\nonumber\\
		C_3&=\frac{-\eta\textsl{g}\Delta_{e1}}{[\Delta_{e1}\Delta_{c1}-\textsl{g}^2][\overline{\Delta}_{c}(\overline{\Delta}_{c}+\overline{\Delta}_{e})-\textsl{g}^2]},\nonumber\\
		C_4&=\frac{-\eta\textsl{g}(2\Delta_{c2}+\Delta_{e1})}{[\Delta_{e2}\Delta_{c2}-\textsl{g}^2][\overline{\Delta}_{c}(\overline{\Delta}_{c}+\overline{\Delta}_{e})-\textsl{g}^2]},\nonumber\\
		C_5&=\frac{\Delta_{e1}}{\Delta_{e1}\Delta_{c1}-\textsl{g}^2},\  C_6=\frac{-\eta\textsl{g}}{\Delta_{e2}\Delta_{c2}-\textsl{g}^2},\label{eq41}
	\end{align}
	where $\overline{\Delta}_x=(\Delta_{x1}+\Delta_{x2})/2$. Notably, we find that $\tilde{g}_{ce}^{(2)}(t)$ is a periodic function about $t$ with the minimum positive period $T=2\pi/\abs{\delta}$, and, if we want to verify the correctness, we need to select an appropriate reference point, i.e., $\tilde{g}_{ce}^{(2)}(0)=\text{Tr}[a^{\dagger2}a^2\rho(t_0)]/\text{Tr}[a^{\dagger}a\rho(t_0)]^2$, where $t_0\gg1$.

	\begin{figure*}[t]
		\includegraphics[width=17.8cm]{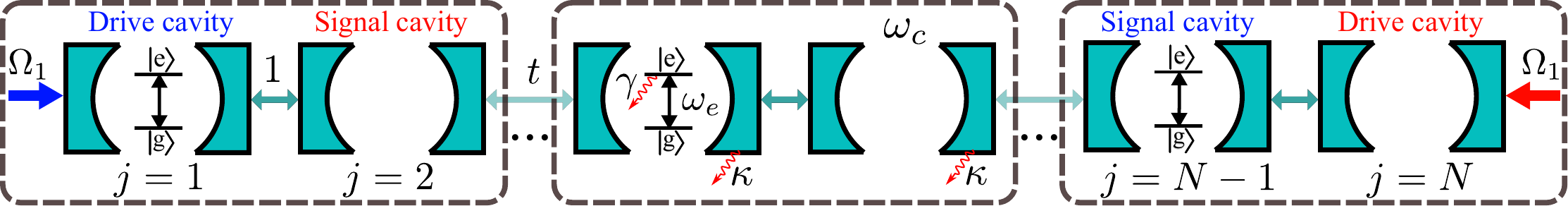}\\
		\caption{Schematic of a dimer JC chain lattice. The first (last) cavity is coherently driven with driving strength $\Omega_1$, and the penultimate (second) cavity is the signal cavity. Each cavity has resonance frequency $\omega_c$ and decay rate $\kappa$, and each odd cavity couples to a two-level atom with transition frequency $\omega_e$ and decay rate $\gamma$. Dashed boxes represent the unit cells; the intracell and intercell couplings strength are $1$ and $t$, respectively.
		}\label{fig5}
	\end{figure*}
	
	As depicted in Fig.~\ref{fig4}(a), when both the cavity and atom are driven by two external fields simultaneously, the second-order ETCF exhibits two minimum points, also known as photon blockade points, compared to the case when only the cavity is driven. This phenomenon occurs due to different input channels creating destructive interference, resulting in the photon blockade effect. In Fig.~\ref{fig4}(c), we find that the $n$-photon antibunching (in $\omega_c-\omega_d=-1.3\kappa$) and bunching (in $\omega_c-\omega_d=0.81\kappa$) effects are enhanced dramatically as $n$ increases. Besides, for the other strong $n$-photon antibunching point, their corresponding detuning almost satisfies $\omega_c-\omega_d=[3+1.22\times(n-3)]\kappa$, where $n=3,4,5$. For different driving frequencies, as shown in Fig.~\ref{fig4}(b), we obtain a dynamical photon effect, and it goes through the cycles of bunching and antibunching effects over time, which the period is $T=2\pi/\abs{\omega_1-\omega_2}=4/\kappa$. In Figs.~\ref{fig4}(b) and \ref{fig4}(c), our analytical solution agrees perfectly with the QUTIP simulations, thereby validating the approach. Meanwhile, we also analyze the problem: How much drive strength can we accept for our method? And give the boundary of driving strength to a certain extent, as shown in Fig.~\ref{fig4}(d). As a result, the case of multiple input channels can trigger the photon blockade; therefore, we could regard the case as a different way to achieve photon blockade. More importantly, the scheme is more feasible for experimental implementations.
	
	No doubt, for the case of multi-atom, i.e., Tavis-Cummings model \cite{TCmodel}, the method is also powerfully effective, and some concrete details solving the second-order ETCF have been presented in Ref.\,\cite{SM13}.
	
	\begin{figure}
		\includegraphics[width=8.6cm]{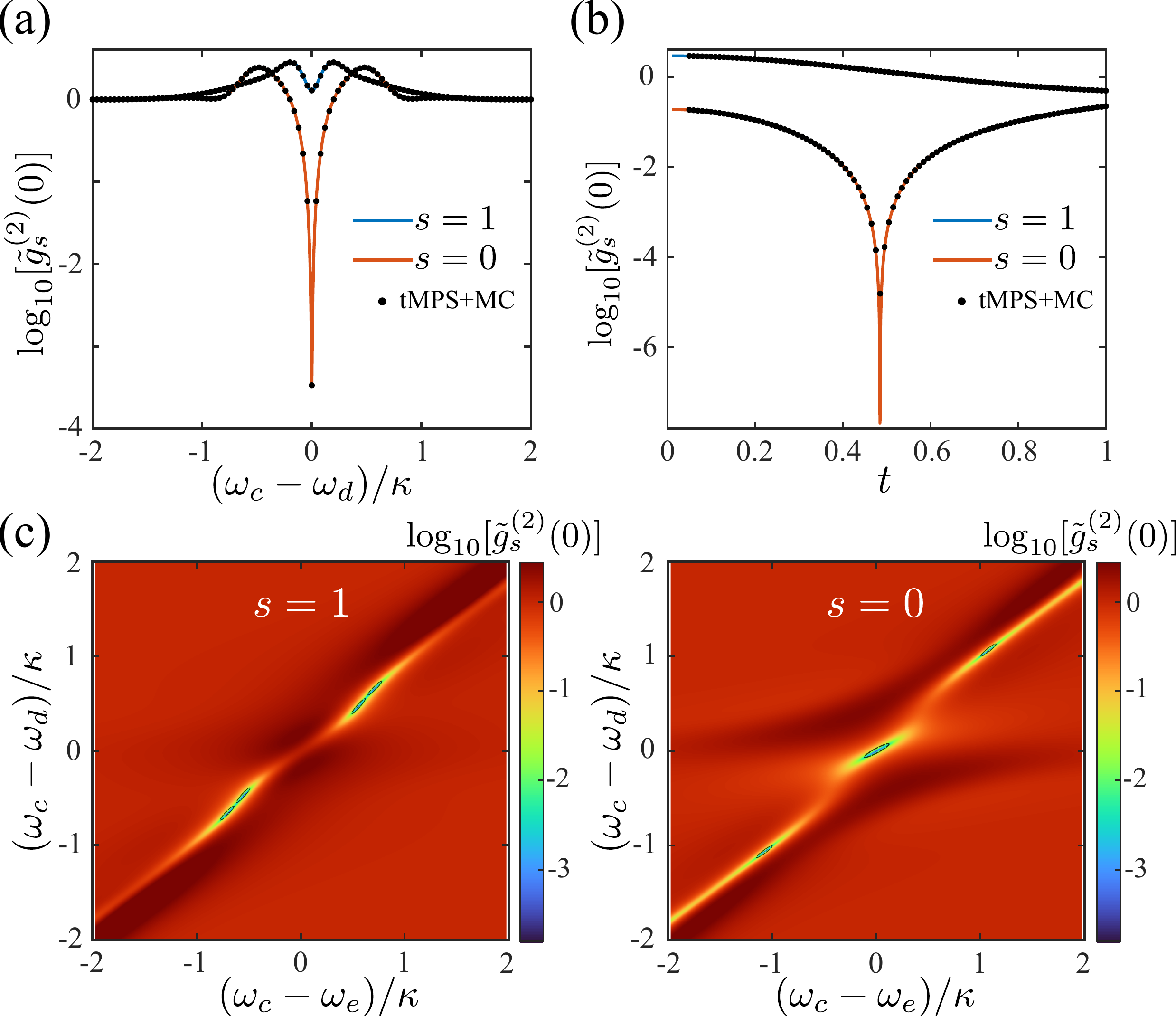}\\
		\caption{Equal-time second-order correlation $\log_{10}[\tilde{g}^{(2)}_s(0)]$ versus the detuning $(\omega_c-\omega_d)/\kappa$ in $\left({\rm{a}}\right)$, the inter-cell hopping $t$ in $\left({\rm{b}}\right)$, and the detuning $(\omega_c-\omega_d)/\kappa$ and $(\omega_c-\omega_e)/\kappa$ in $\left({\rm{c}}\right)$. Here, the black dots in (a) and (b) represent the numerical comparison by using time-evolved matrix product state (tMPS) and Monte Carlo (MC). The black dashed circles in $\left({\rm{c}}\right)$ highlight the minimum point of the correlation function. In all subplots, the system parameters are given by $N_c=8$, $\kappa=1$, $\gamma=0.8\kappa$, and $\textsl{g}=0.6\kappa$. Specially, $\omega_c=\omega_e$ for $\left({\rm{a}}\right)$, $\omega_c=\omega_e=\omega_d$ for $\left({\rm{b}}\right)$, and $t=0.5$ for (a) and (c). 
		}\label{fig6}
	\end{figure}
	\subsection{Coupled Cavity Array Model}\label{VIB}
	This section considers a dimer JC chain system, as shown in Fig.~\ref{fig5}. The Hamiltonian is given by
	\begin{align}
		H_{\text{CA}}^{(s)}&=\omega_c\sum_{j=1}^{N_c}a_j^\dagger a_j+\sum_{j=1}^{Nc/2}[\omega_e\sigma_{j}^\dagger\sigma_{j}+\textsl{g}(a_{2j-s}^\dagger\sigma_{j}+\text{H.c.})]\nonumber\\ &+[\sum_{j=1}^{N_c/2}a_{2j-1}^\dagger a_{2j}+t\sum_{j=1}^{N_c/2-1}a_{2j}^\dagger a_{2j+1}+\text{H.c.}].\label{eq42}
	\end{align}
	Here, $a_j^{(\dagger)}$ and $\omega_c$ are the annihilation (creation) operator and resonant frequency of the $j$-th cavity mode; $\sigma_j$, $\omega_e$, and $\textsl{g}$ are the lowering operator, transition frequency, and cavity coupling strength of the $j$-th atom. Here, $t$ represents the inter-cell hopping (with the intra-cell hopping normalized to unity), the number of the cavity, $N_c$, is even, and $s=0(1)$ denotes only the even (odd) cavity coupled to the atom. The Hamiltonian describes a Su–Schrieffer–Heeger (SSH) model \cite{PhysRevLett.42.1698}, when the system does not contain atoms. 
	
	Meanwhile, the Lindblad master equation is 
	\begin{align}
		\dot{\rho}=-i[H^{(s)}_{\text{tot}},\rho]+
		\kappa\sum_{j=1}^{Nc}\mathcal{D}[a_j]\rho+\gamma\sum_{j=1}^{Nc/2}\mathcal{D}[\sigma_j]\rho,\label{eq43}
	\end{align} 
	where $\kappa$ and $\gamma$ are the cavity and atomic decay rates, respectively, and $H_{\text{tot}}^{(s)}=H_{CA}^{(1)}+H_{\text{d}}{(s)}$. The driving term is $H_{\text{d}}^{(s)}= [\Omega_1^*(a_1\delta_{s,1}+a_N\delta_{s,0})e^{i\omega_dt}+\text{H.c.}]$, and $s=1(0)$ correspond to red (blue) arrow within Fig.~\ref{fig5}. Next, we will mainly study the impact of the different positions of atoms within an unit cell on the statistical property of output light from a signal cavity, i.e., $s=0$ and $s=1$. Note that we will not provide the derivation steps or calculating procedures, all the data results come from the QCS code \cite{Qcs}.
	
	\begin{figure*}
		\includegraphics[width=17.8cm]{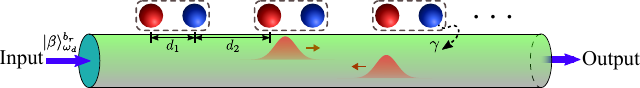}\\
		\caption{Schematic of a dimer atom chain coupled to a 1D waveguide. $\gamma$ represents the decay rates of atom into the 1D waveguide. The input state is described by $|\beta\rangle_{\omega_d}^{b_r}$. Dashed boxes indicate the unit cells consisting of red and blue balls; the intra-cell distance is $d_1$ and the inter-cell atoms are offset by the distance $d_2$.
		}\label{fig7}
	\end{figure*}
	
	As shown in Fig.~\ref{fig6}(a), there is an excellent distinction between $s=1$ and $0$ in resonance point ($\omega_c=\omega_e=\omega_d$), and this indicates that different configurations (i.e., the atoms only located in odd or even cavity) could induce the strong photon blockade effect. Meanwhile, in resonance point, we also study the impact of inter-cell hopping $t$ on the second-order ETCF with different configurations, as shown in Fig.~\ref{fig6}(b). Notably, for the case of atoms located at the even cavity ($s=0$), there is a strong antibunching point at $t\approx0.48$, while this point disappears for the other case ($s=1$). Besides, when we tune the detuning of cavity and atom frequencies ($\omega_c-\omega_e\neq0$), the two configurations both create photon blockade effect, as shown in Fig.~\ref{fig6}(c). In Fig.~\ref{fig6}(c), the transition from $s=1$ to $0$ seems like two small antibunching regions in $s=1$ merging into a larger antibunching region near the resonance position in $s=0$. Specifically, for $s=1$, the innermost and outermost two strong antibunching regions along the axis $\omega_e-\omega_d=0$ move toward the resonance point and the non-resonance point ($\omega_c-\omega_e=\pm\kappa$), respectively, and finally forms the case of $s=0$.
	
	Finally, we use a complicatedly complex numerical method (tMPS and MC) to check out our analytical expression, and the method agrees perfectly with the analytical solution except for the minimum point. The error in strong antibunching points mainly comes from a finite driving strength ($\Omega_1=0.01\kappa$) in numerical simulations and truncation errors in the tMPS. Nevertheless, the compactly analytical solution is still highly effective and correct.
	\subsection{Spin-1/2 Systems Coupled To A 1D Waveguide Model}\label{VIC}
	In this section, we consider a waveguide QED system, i.e., a dimer atom chain side-coupled to a 1D waveguide, as shown in Fig.~\ref{fig7}. The Hamiltonian for the atom chain reads $H_{\text{sys}}=\omega_e\sum_{j=1}^N\sigma_j^\dagger\sigma_j$, where $\omega_e$ is the transition frequency of atom. Meanwhile, the Hamiltonian of a 1D waveguide is
	\begin{align}\label{eq44}
		H_{\text{wg}}=\sum_{\mu=l,r}\int\dd\omega\ \omega b_{\mu}^\dagger(\omega)b_{\mu}(\omega),
	\end{align}
	where the $b_\mu(\omega)$ are bosonic annihilation operators for the right ($\mu=r$) and left ($\mu=l$) moving waveguide modes of frequency $\omega$. Note from Eq.~(\ref{eq44}) that we implicitly assumed a linear dispersion relation for the degrees of freedom of the waveguide. The system-waveguide interaction can be characterized by the rotating wave approximation Hamiltonian
	\begin{align}\label{eq45}
		H_{\text{int}}=i\sum_{j=1}^N\sum_{\mu=l,r}\int\dd\omega
		\sqrt{\frac{\gamma_\mu}{2\pi}}\left[b_\mu^\dagger(\omega)\sigma_{j} e^{-i\omega x_j/v_\mu}-\text{H.c.}\right],
	\end{align}
	where $\gamma_l$ and $\gamma_r$ are the decay rates into to the left- ($v_l<0$) and right- ($v_r>0$) moving waveguide modes, respectively, $v_\mu$ denotes the corresponding group velocities, and $x_j$ represents the position of the $j$-th atom along the waveguide, i.e., $x_{2j+1}=j(d_1+d_2)$ and $x_{2j}=j(d_1+d_2)-d_2$. 
	
	\begin{figure}
		\includegraphics[width=8.6cm]{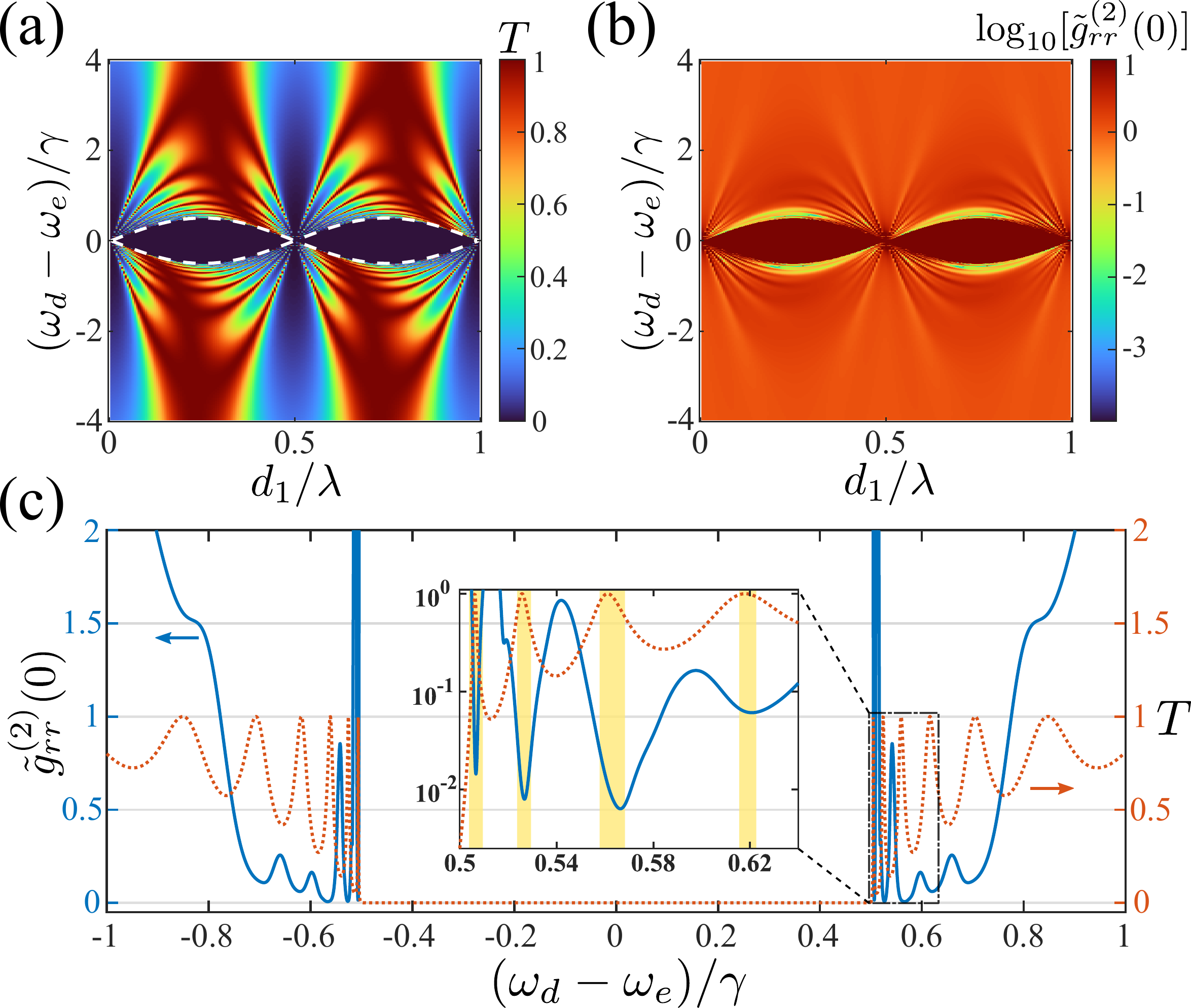}\\
		\caption{The single-photon transmission $T$ for $\left({\rm{a}}\right)$ and the second-order ETCF $\log_{10}[\tilde{g}_{rr}^{(2)}(0)]$ for $\left({\rm{b}}\right)$ versus the detuning $(\omega_d-\omega_e)/\gamma$ and the distance $d_1/\lambda$. In $\left({\rm{a}}\right)$, the boundary of $T=0$ corresponds to the white dashed line. In $\left({\rm{b}}\right)$, to distinguish the local minima more clearly, all regions for which $\tilde{g}_{rr}^{(2)}(0)>10$ are colored dark red. $\left({\rm{c}}\right)$ The second-order ETCF $\tilde{g}_{rr}^{(2)}(0)$ (left-hand  longitudinal axis) and single-photon transmission $T$ (right-hand longitudinal axis) versus the detuning $(\omega_d-\omega_e)/\gamma$. The yellow regions in the inset indicate a close correspondence between high transmissivity ($T\approx 1$) and strong antibunching effect [$\tilde{g}_{rr}^{(2)}(0)\ll 1$]. In all subplots, the system parameters are given by $N=20$, $d_1+d_2=\lambda$, and $\lambda=2\pi/k$. Especially $d_1/\lambda=\frac{1}{4}$ for $\left({\rm{c}}\right)$.
		}\label{fig8}
	\end{figure}
	
	First, we assume $\gamma_l=\gamma_r=\gamma/2$, $v_r=-v_l=v$, and $k=\omega_e/v$. Then, we consider an incoming coherent state $|\beta\rangle_{\omega_d}^{b_r}$, and the coherent amplitude is small enough, i.e., $\abs{\beta}\to0$. Finally, following the standard procedures \cite{PhysRevA.91.042116,PhysRevA.95.063809} (such as the usual Born-Markov and secular approximations), the Lindblad master equation reads as
	\begin{align}
		\dot{\rho}&=-i[H_{\text{sys}}+H_{\text{coh}}+H_{\text{d}},\rho]\nonumber\\
		&+\gamma\sum_{i,j}\cos(k\abs{x_i-x_j})\left(\sigma_j\rho\sigma_i^\dagger-\frac{1}{2}\{\sigma_i^\dagger\sigma_j,\rho\}\right),\label{eq46}
	\end{align}
	where $H_{\text{coh}}=(\gamma/2)\sum_{i,j}\sin(k\abs{x_i-x_j})\sigma_j^\dagger\sigma_i$, and $H_{\text{d}}=\sum_{j=1}^N[\beta^*\sqrt{\gamma/4\pi} \exp(i\omega_dt-ikx_j)\sigma_j+\text{H.c.}]$. 
	
	Next, we will study the statistical properties of photons from the output channel $b_r$, such as the single-photon transmission and the second-order ETCF. Like the second example, all data results come from the QCS code \cite{Qcs}.
	
	When $d_1+d_2=2\pi/k$, $H_{\text{coh}}$ in single-excitation subspace has a chiral symmetry ensuring its spectrum is symmetric about $0$. The spectrum is
	\begin{align}\label{eq47}
		E_m=\pm\abs{\frac{\gamma\sin(kd_1)}{1-\exp(2i\pi m/N)}},m=1,3,\ldots,N-1.
	\end{align}
	Notably, when the frequency $\omega_d$ of incoming photons is within the range, i.e., $[-\min\abs{E_m}+\omega_e,\min\abs{E_m}+\omega_e]$, the transmission is almost zero, and also is accompanied by a pronounced bunching effect, as shown in Fig.~\ref{fig8}(a) and \ref{fig8}(b). And, the white dashed corresponds to $\pm \min\abs{E_m}$. Meanwhile, when we choose a maximal window corresponding to $T\approx0$, i.e., $kd_1=\pi/2$, as shown in Fig.~\ref{fig8}(c), we find that the transmission happens quickly oscillation near the $\abs{\omega_d-\omega_e}\gtrsim \min\abs{E_m}$, and the peaks of oscillation are equal to one ($T=1$). The first four peaks also are accompanied by a strong antibunching effect, respectively, as shown in the yellow regions within Fig.~\ref{fig8}(c).
	\section{Conclusions And Outlook}
	To sum up, we introduce a new physical quantity of the probability amplitude with equal-time probing multiple photons, and the probability amplitude has a compactly analytical expression for any quantum systems satisfying the U(1) symmetry. Based on this analytical expression, we could simply obtain an $n$th-order ETCF under a weak coherent drive, because the correlation function completely depends on the probability amplitude. In addition, we also discuss the classification of input and output channels, and we correspondingly give the probability amplitude and correlation function. We find that multiple input channels could bring up an interference effect, and the phenomenon is beneficial to study photon blockades for us. Moreover, multiple output channels also can help us to study the cross correlation. We also prove that our analytical expression works for identical input and output channels, providing excellent convenience for studying the waveguide QED systems. Finally, we consider three examples in order to illustrate the advantages of our method. Not only do we find that multiple input channels can trigger the photon blockade effect compared to a single input channel, but we can also apply our method to solve large-dimensional systems, such as multi-cavity and multi-atom models. Crucially, the probability amplitude can also be used to calculate other important physical quantities, such as transmission spectrum, reflectance spectrum, and multi-bundle correlation functions \cite{munoz_emitters_2014, PhysRevLett.117.203602}. We also provide a user-friendly open-source library in Python and expect our method here to be further applied and extended.
	\section{ACKNOWLEDGMENTS}
	We thank Professor T.~Shi for valuable discussions. This work is supported by the National Key Research and Development Program of China Grant No. 2021YFA1400700 and the National Science Foundation of China Grant No. 11974125. C.S. acknowledges the financial support by the China Scholarship Council and the Japanese Government (Monbukagakusho-MEXT) Scholarship under Grant No.~211501 and the RIKEN Junior Research Associate Program.
	\begin{widetext}
		\appendix
		\section{The derivations of input-output relations and scattering matrix elements}\label{A}
		For the total Hamiltonian~(\ref{eq1}), the Heisenberg equations of motion for the operators $b_i(\omega)$ and $c_j(\omega)$ are
		\begin{align}\label{motion}
			i\frac{\dd b_i(\omega)}{\dd t}=[b_i(\omega),H_{\text{tot}}]=\omega b_i(\omega)+\xi_{b,i}o_i,\ 
			i\frac{\dd c_j(\omega)}{\dd t}=[c_j(\omega),H_{\text{tot}}]=\omega c_j(\omega)+\xi_{c,j}o_j.
		\end{align}
		In the assumptions $t>t_0$ and $t_0\to-\infty$, the differential equations ~(\ref{motion}) can be solved:
		\begin{align}\label{motion1}
			b_i(\omega,t)&=b_i(\omega,t_0)e^{-i\omega(t-t_0)}-i\xi_{b,i}\int_{t_0}^{t}o_i(\tau)e^{-i\omega(t-\tau)}\dd\tau,\nonumber\\
			c_j(\omega,t)&=c_j(\omega,t_0)e^{-i\omega(t-t_0)}-i\xi_{c,j}\int_{t_0}^{t}o_j(\tau)e^{-j\omega(t-\tau)}\dd\tau.
		\end{align}
		We then integrate ~(\ref{motion1}) with respect to $\omega$ to obtain
		\begin{align}
			\Phi_b(t)&=\frac{1}{\sqrt{2\pi}}\int_{-\infty}^{+\infty}b_i(\omega,t)\dd\omega=b_{i,\text{in}}(t)-\frac{i\xi_{b,i}}{\sqrt{2\pi}}\int_{t_0}^{t}o_i(\tau)\int_{-\infty}^{+\infty}e^{-i\omega(t-\tau)}\dd\omega \dd\tau=b_{i,\text{in}}(t)-\frac{i\sqrt{\kappa_{b,i}}}{2}o_i(t),\nonumber\\
			\Phi_c(t)&=\frac{1}{\sqrt{2\pi}}\int_{-\infty}^{+\infty}c_j(\omega,t)\dd\omega=c_{j,\text{in}}(t)-\frac{i\xi_{c,j}}{\sqrt{2\pi}}\int_{t_0}^{t}o_j(\tau)\int_{-\infty}^{+\infty}e^{-i\omega(t-\tau)}\dd\omega \dd\tau=c_{j,\text{in}}(t)-\frac{i\sqrt{\kappa_{c,j}}}{2}o_j(t),\label{Phi1}
		\end{align}
		with the input operators
		\begin{align}\label{in}
			b_{i,\text{in}}(t)=\frac{1}{\sqrt{2\pi}}\int_{-\infty}^{+\infty}b_i(\omega,t_0)e^{-i\omega(t-t_0)}\dd\omega,\quad
			c_{j,\text{in}}(t)=\frac{1}{\sqrt{2\pi}}\int_{-\infty}^{+\infty}c_j(\omega,t_0)e^{-i\omega(t-t_0)}\dd\omega.
		\end{align}
		Similarly, in the assumptions $t_1>t$ and $t_1\to+\infty$, Eq.~(\ref{motion}) can be written as
		\begin{align}\label{motion5}
			b_i(\omega,t)&=b_i(\omega,t_1)e^{-i\omega(t-t_1)}+i\xi_{b,i}\int_{t_1}^{t}o_i(\tau)e^{-i\omega(t-\tau)}\dd\tau,\nonumber\\
			c_j(\omega,t)&=c_j(\omega,t_1)e^{-i\omega(t-t_1)}+i\xi_{c,j}\int_{t_1}^{t}o_j(\tau)e^{-j\omega(t-\tau)}\dd\tau.
		\end{align}
		We integrate ~(\ref{motion5}) concerning $\omega$ to obtain
		\begin{align}
			\Phi_b(t)&=\frac{1}{\sqrt{2\pi}}\int_{-\infty}^{+\infty}b_i(\omega,t)\dd\omega=b_{i,\text{out}}(t)+\frac{i\xi_{b,i}}{\sqrt{2\pi}}\int_{t_1}^{t}o_i(\tau)\int_{-\infty}^{+\infty}e^{-i\omega(t-\tau)}\dd\omega \dd\tau=b_{i,\text{out}}(t)+\frac{i\sqrt{\kappa_{b,i}}}{2}o_i(t),\nonumber\\
			\Phi_c(t)&=\frac{1}{\sqrt{2\pi}}\int_{-\infty}^{+\infty}c_j(\omega,t)\dd\omega=c_{j,\text{out}}(t)+\frac{i\xi_{c,j}}{\sqrt{2\pi}}\int_{t_1}^{t}o_j(\tau)\int_{-\infty}^{+\infty}e^{-i\omega(t-\tau)}\dd\omega \dd\tau=c_{j,\text{out}}(t)+\frac{i\sqrt{\kappa_{c,j}}}{2}o_j(t),\label{Phi2}
		\end{align}
		with the output operators
		\begin{align}\label{out}
			b_{i,\text{out}}(t)=\frac{1}{\sqrt{2\pi}}\int_{-\infty}^{+\infty}b_i(\omega,t_1)e^{-i\omega(t-t_1)}\dd\omega,\quad
			c_{j,\text{out}}(t)=\frac{1}{\sqrt{2\pi}}\int_{-\infty}^{+\infty}c_j(\omega,t_1)e^{-i\omega(t-t_1)}\dd\omega.
		\end{align}
		Combining with Eqs.~(\ref{Phi1}) and (\ref{Phi2}), we get the input-output relations:
		\begin{align}\label{A8}
			b_{i,\text{out}}(t)=b_{i,\text{in}}(t)-i\sqrt{\kappa_{b,i}}o_i(t),\ 
			c_{j,\text{out}}(t)=c_{j,\text{in}}(t)-i\sqrt{\kappa_{c,j}}o_j(t).
		\end{align}
		
		Due to the property of Møller wave operators, i.e., $\Omega_-^+ \Omega_-=\Omega_+^\dagger\Omega_+=I$, we have
		\begin{align}
			S^{\mu\nu}_{p_1\ldots p_n;k_1\ldots k_n}&=\langle0|[\prod_{l=1}^{n}\nu_l(p_l)]\hat{S}[\prod_{l=1}^{n}\mu_l^\dagger(k_l)]|0\rangle=\langle0|\Omega_-^\dagger[\prod_{l=1}^{n}\Omega_-\nu_l(p_l)\Omega_-^\dagger]\Omega_-\hat{S}\Omega_+^\dagger[\prod_{l=1}^{n}\Omega_+\mu_l^\dagger(k_l)\Omega_+^\dagger]\Omega_+|0\rangle\nonumber\\
			&=\langle0|[\prod_{l=1}^{n}\Omega_-\nu_l(p_l)\Omega_-^\dagger]\Omega_-\Omega_-^\dagger\Omega_+\Omega_+^\dagger[\prod_{l=1}^{n}\Omega_+\mu_l^\dagger(k_l)\Omega_+^\dagger]|0\rangle\nonumber\\
			&=\langle0|[\prod_{l=1}^{n}\Omega_-\nu_l(p_l)\Omega_-^\dagger][\prod_{l=1}^{n}\Omega_+\mu_l^\dagger(k_l)\Omega_+^\dagger]|0\rangle\nonumber\\
			&=\langle0|[\prod_{l=1}^{n}\nu_{l,\text{out}}(p_l)][\prod_{l=1}^{n}\mu_{l,\text{in}}^\dagger(k_l)]|0\rangle,
		\end{align}
		where 
		\begin{align} 
			\mu_{l,\text{in}}(k_l)&\equiv\Omega_+\mu_l(k_l)\Omega_+^\dagger=e^{iH_{\text{tot}}t_0}e^{-iH_{\text{B}}t_0}\mu_l(k_l)e^{iH_{\text{B}}t_0}e^{-iH_{\text{tot}}t_0},\label{definition_in}\\
			\nu_{l,\text{out}}(p_l)&\equiv\Omega_-\nu_l(p_l)\Omega_-^\dagger=e^{iH_{\text{tot}}t_1}e^{-iH_{\text{B}}t_1}\nu_l(p_l)e^{iH_{\text{B}}t_1}e^{-iH_{\text{tot}}t_1}\label{definition_out}.
		\end{align}
		According to Eq.~(\ref{in}) and Eq.~(\ref{out}), for the operator $b_{i,\text{in}}(t)$ we have
		\begin{align}
			b_{i,\text{in}}(t)&=\frac{1}{\sqrt{2\pi}}\int_{-\infty}^{+\infty}b_i(\omega,t_0)e^{-i\omega(t-t_0)}\dd\omega=\frac{1}{\sqrt{2\pi}}\int_{-\infty}^{+\infty}e^{iH_{\text{tot}}t_0}b_i(\omega)e^{i\omega t_0}e^{-iH_{\text{tot}}t_0}e^{-i\omega t}\dd\omega\nonumber\\
			&=\frac{1}{\sqrt{2\pi}}\int_{-\infty}^{+\infty}e^{iH_{\text{tot}}t_0}e^{-iH_{\text{B}} t_0}b_i(\omega)e^{iH_{\text{B}} t_0}e^{-iH_{\text{tot}}t_0}e^{-i\omega t}\dd\omega\nonumber\\
			&=\frac{1}{\sqrt{2\pi}}\int_{-\infty}^{+\infty}b_{i,\text{in}}(\omega)e^{-i\omega t}\dd\omega,
		\end{align}
		which automatically satisfy the inverse Fourier transform by the definition~(\ref{definition_in}), and the same is true for $c_{j,\text{out}}(t)$, i.e., $c_{j,\text{out}}(t)=\mathscr{F}^{-1}[c_{j,\text{out}}(\omega)]$, where $\mathscr{F}$ represent the Fourier transform. Note that in the second line, we took advantage of the fact that $[H_{\text{B}}, b_i(\omega)]=-\omega b_i(\omega)$. As a result, the definitions ~(\ref{definition_in}) and ~(\ref{definition_out}) are self-consistent.  
		
		\section{The derivation of Eq.~(\ref{eq6})}\label{B}
		In order to demonstrate the equivalence between the scattering matrix and master equation methods, we only need to calculate the numerator of Eq.~(\ref{eq6}), and the computations are as below:
		\begin{align}
			\langle\psi_{\text{out}}|c_j^{\dagger n}(t)  c_j^n(t)|\psi_{\text{out}}\rangle&=\langle\psi_{\text{in}}|\hat{S}^\dagger\Omega_-^\dagger [\Omega_-c_j^\dagger(t)\Omega_-^\dagger]^n[\Omega_-c_j(t)\Omega_-^\dagger]^n\Omega_-\hat{S}|\psi_{\text{in}}\rangle
			=\langle\psi_{\text{in}}|\Omega_+^\dagger c^{\dagger n}_{\text{out},j}(t)c^n_{\text{out},j}(t)\Omega_+|\psi_{\text{in}}\rangle\nonumber\\
			&=\langle\psi_{\text{in}}|\Omega_+^\dagger [c^\dagger_{\text{in},j}(t)+i\sqrt{\kappa_{c,j}}o_j^\dagger(t)]^n[c_{j,\text{in}}(t)-i\sqrt{\kappa_{c,j}}o_j(t)]^n\Omega_+|\psi_{\text{in}}\rangle\label{numerator}
		\end{align}
		Then, according to these definitions of $\Omega_+$, $|\psi_{\text{in}}\rangle$ and $c_{j,\text{in}}(t)$, we have
		\begin{align}
			c_{j,\text{in}}(t)\Omega_+|\psi_{\text{in}}\rangle&=[\int_{-\infty}^{+\infty}e^{iH_{\text{tot}}t_0}e^{-iH_{\text{B}}t_0}c_j(\omega)e^{iH_{\text{B}}t_0}e^{-iH_{\text{tot}}t_0}e^{-i\omega t}\dd\omega][e^{iH_{\text{tot}}t_0}e^{-iH_{\text{B}}t_0}][\mathscr{N}|\beta\rangle_{\omega_d}^{b_i}\otimes|0\rangle_{\text{B}}\otimes|g\rangle_{\text{s}}] \nonumber\\
			&=\int_{-\infty}^{+\infty}e^{iH_{\text{tot}}t_0}e^{-iH_{\text{B}}t_0}c_j(\omega)e^{-i\omega t}\mathscr{N}|\beta\rangle_{\omega_d}^{b_i}\otimes|0\rangle_{\text{B}}\otimes|g\rangle_{\text{s}}\ \dd\omega \nonumber\\
			&=0=\langle\psi_{\text{in}}|\Omega_+^\dagger c^\dagger_{\text{in},j}(t)\label{zero_condition},
		\end{align}
		where $\mathscr{N}$ is a normalization factor, and the quantum causality condition about $o_j(t)$ and $o_j^\dagger(t)$ at equal time is
		\begin{align}\label{causality}
			[o_j(t),c_{j,\text{in}}(t)]=[o^\dagger_j(t),c_{j,\text{in}}(t)]=[o_j(t),c^\dagger_{\text{in},j}(t)]=[o^\dagger_j(t),c^\dagger_{\text{in},j}(t)]=0.
		\end{align}
		Hence, combining with Eqs.~(\ref{zero_condition}) and (\ref{causality}), (\ref{numerator}) can be further simplified as
		\begin{align}
			\langle\psi_{\text{out}}|c_j^{\dagger n}(t)  c_j^n(t)|\psi_{\text{out}}\rangle&=\kappa_{c,j}^n\langle\psi_{\text{in}}|e^{iH_{\text{B}}t_0}e^{-iH_{\text{tot}}t_0}o_j^{\dagger n}(t)o_j^n(t)e^{iH_{\text{tot}}t_0}e^{-iH_{\text{B}}t_0}|\psi_{\text{in}}\rangle\nonumber\\
			&=\kappa_{c,j}^n\text{Tr}[o_j^{\dagger n}(t)o_j^n(t)e^{iH_{\text{tot}}t_0}e^{-iH_{\text{B}}t_0}|\psi_{\text{in}}\rangle\langle\psi_{\text{in}}|e^{iH_{\text{B}}t_0}e^{-iH_{\text{tot}}t_0}]\nonumber\\
			&=\kappa_{c,j}^n\text{Tr}[o_j^{\dagger n}o_j^ne^{-iH_{\text{tot}}(t_{\infty}+t)}\rho(0)e^{iH_{\text{tot}}(t_{\infty}+t)}]\nonumber\\
			&=\kappa_{c,j}^n\text{Tr}[o_j^{\dagger n}o_j^nU(t_\infty+t,0)\rho(0)U^\dagger(t_\infty+t,0)]\nonumber\\
			&=\kappa_{c,j}^n\text{Tr}_s[o_j^{\dagger n}o_j^n\rho_s(t_\infty+t)]\label{D4},
		\end{align}
		where $\rho_s(t_\infty+t)=\text{Tr}_\text{B}[U(t_\infty+t,0)\rho(0)U^\dagger(t_\infty+t,0)]$, $\rho(0)=e^{-iH_\text{B}t_0}|\psi_{\text{in}}\rangle\langle\psi_{\text{in}}|e^{iH_\text{B}t_0}$, $t_0=-\infty=-t_\infty$, and $\text{Tr}_s$ and $\text{Tr}_\text{B}$, respectively, represent partial trace for systems and baths. For the sake of conciseness, we introduce a time-dependent displacement operator 
		\begin{align}
			\mathcal{D}_t\{\beta(\omega)\}=\exp(\int_{-\infty}^{\infty}\dd\omega\left[\beta(\omega)e^{-i\omega t}b_i^\dagger(\omega)-\beta^*(\omega)e^{i\omega t}b_i(\omega)\right]),
		\end{align}
		where $\beta(\omega)=\beta\delta(\omega-\omega_d)$ when a local system be coherently driven, and the free evolution of input state can be also rewritten as $e^{-iH_\text{B}t_0}|\psi_{\text{in}}\rangle=\mathcal{D}_{0}\{\beta(\omega)e^{-i\omega t_0}\}|0\rangle$. Finally, in order to be consistent with Eq.~(\ref{eq4}), we need to demonstrate the equation:
		\begin{align}
			\rho_s(t)&=\text{Tr}_{\text{B}}[\mathcal{D}^\dagger_t\{\beta(\omega)e^{-i\omega t_0}\}U(t,0)\mathcal{D}_{0}\{\beta(\omega)e^{-i\omega t_0}\}|0\rangle\langle0|\mathcal{D}^\dagger_{0}\{\beta(\omega)e^{-i\omega t_0}\}U^\dagger(t,0)\mathcal{D}_t\{\beta(\omega)e^{-i\omega t_0}\}]\nonumber\\
			&=\text{Tr}_{\text{B}}[\widetilde{U}(t,0)\widetilde{\rho}(0)\widetilde{U}^\dagger(t,0)]=\widetilde{\rho}_s(t),
		\end{align}
		where $\widetilde{\rho}(0)=|0\rangle\langle0|$, and $|0\rangle$ represent the vacuum state of the total system. And beyond that, according to the Mollow transformation, the new evolution operator $\widetilde{U}(t,0)$ has
		\begin{align}
			i\frac{d}{dt}\widetilde{U}(t,0)=\widetilde{H}(t)\widetilde{U}(t,0)\quad \text{with}\quad \widetilde{H}(t)=H_{\text{tot}}+H_{\text{d}}^i(t+t_0).
		\end{align}
		Therefore, following the standard procedures \cite{Pathria1996,lidar_lecture_2020} to trace out the heat baths degrees of freedom and applying the Born-Markov approximation, 
		the density matrix $\widetilde{\rho}_s(t)$ satisfies the master equation
		\begin{align}\label{master_equation}
			\partial_t\widetilde{\rho}_s(t)=-i[H_{\text{sys}}\{o_k\}+H_{\text{d}}^i(t),\widetilde{\rho}_s(t)]+\sum_\alpha\gamma_\alpha\left(L_\alpha\widetilde{\rho}_s(t)L_\alpha^\dagger-\frac{1}{2}\{L_\alpha^\dagger L_\alpha,\widetilde{\rho}_s(t)\}\right)\equiv\mathcal{L}\widetilde{\rho}_s(t)
		\end{align}
		with Liouvillian operator $\mathcal{L}$. Equation (\ref{master_equation}) actually is Eq.~(\ref{eq4}), so Eq.~(\ref{D4}) can be further written as 
		\begin{align}\label{D9}
			\langle\psi_{\text{out}}|c_j^{\dagger n}(t)  c_j^n(t)|\psi_{\text{out}}\rangle=\kappa_{c,j}^n\text{Tr}_s[o_j^{\dagger n}o_j^n\rho_s(t_\infty+t)]=\kappa_{c,j}^n\text{Tr}_s[o_j^{\dagger n}o_j^n\widetilde{\rho}_s(t_\infty+t)].
		\end{align}
		By plugging the Eq.~(\ref{D9}) into (\ref{eq6}), we have
		\begin{align}
			g_{jj}^{(n)}(0)&=\frac{\langle\psi_{\text{out}}|c_j^{\dagger n}(t)  c_j^n(t)|\psi_{\text{out}}\rangle}{\ \ \langle\psi_{\text{out}}| c^\dagger_j(t)c_j(t)|\psi_{\text{out}}\rangle ^n}=\frac{\kappa_{c,j}^n\text{Tr}_s[o_j^{\dagger n}o_j^n\widetilde{\rho}_s(t_\infty+t)]}{\kappa_{c,j}^n\text{Tr}_s[o_j^{\dagger }o_j\widetilde{\rho}_s(t_\infty+t)]^n}=\frac{\text{Tr}_s[o_j^{\dagger n}o_j^n\rho_{\text{ss}}]}{\text{Tr}_s[o_j^{\dagger }o_j\rho_{\text{ss}}]^n},
		\end{align}
		where $\rho_{\text{ss}}$ represents the steady state, i.e., $\rho_{\text{ss}}=\lim\limits_{t\to\infty}\widetilde{\rho}_s(t)=\widetilde{\rho}_s(t_\infty+t)$.

		\section{The derivations of Eqs.~(\ref{eq13}-\ref{eq16})}\label{C}
		According to quantum field theory, the time-ordered $2n$-point Green's function~(\ref{eq12}) can be written as the path integral formulation, i.e.,
		\begin{align}
			G^{\mu\nu}(t^\prime_{B_n};t_{D_n})=(-1)^n\frac{\int\mathcal{D}\left[\left\{b_k(\omega),b_k^*(\omega),c_k(\omega),c_k^*(\omega)\right\},\left\{o_k,o_k^*\right\}\right]\prod_{l=1}^{n}\left[o_{\nu_l}(t^\prime_l)o^{*}_{\mu_l}(t_l)\right]e^{i\int \dd t\mathcal{L}}}{\int\mathcal{D}\left[\left\{b_k(\omega),b_k^*(\omega),c_k(\omega),c_k^*(\omega)\right\},\left\{o_k,o_k^*\right\}\right]e^{i\int \dd t\mathcal{L}}},\label{path1}
		\end{align}
		where $\mathcal{L}$ is the Lagrangian of the total system, and “$\left\{\cdots\right\}$” represents all possible modes within brackets. For the total Hamiltonian~(\ref{eq1}), the $\mathcal{L}$ is
		\begin{align}
			\mathcal{L}=&\sum_k\int\dd\omega [b_k^*(\omega)\left(i\partial_t-\omega\right)b_k(\omega)+c_k^*(\omega)\left(i\partial_t-\omega\right)c_k(\omega)]-\nonumber\\
			&\sum_k\int\dd\omega\left[\xi_{b,k}b_k^*(\omega) o_k+\xi_{c,k}c^*_k(\omega) o_k+\xi^*_{b,k}b_k(\omega) o_k^*+\xi^*_{c,k}c_k(\omega) o^*_k\right]+\mathcal{L}_{\text{sys}},
		\end{align}
		where $\mathcal{L}_{\text{sys}}$ is the system's Lagrangian associating with the Hamiltonian $H_{\text{sys}}\{o_k\}$. For the sake of convenience, we define a symbol, i.e., $A=\partial_t+i\omega$. In order to compute $A^{-1}$, We have
		\begin{align}
			A\varPi_\omega(t-t^\prime)=\delta(t-t^\prime)\Longrightarrow\varPi_\omega(t-t^\prime)=\int \frac{\dd k}{2\pi}e^{-ik(t-t^\prime)}\frac{i}{k-\omega+i0^+}=e^{-i\omega(t-t^\prime)}\theta(t-t^\prime),
		\end{align}
		where the function $\varPi_\omega(t-t^\prime)$ is an inverse of $A$. Hence, we only calculate the functional integration of waveguides in the denominator of Eq.~(\ref{path1}):
		\begin{align}
			&\int\mathcal{D}\left[\left\{b_k(\omega),b_k^*(\omega),c_k(\omega),c_k^*(\omega)\right\}\right]e^{i\int \dd t\mathcal{L}}=e^{i\int \dd t\mathcal{L}_{\text{sys}}}\int\mathcal{D}\left[\left\{b_k(\omega),b_k^*(\omega),c_k(\omega),c_k^*(\omega)\right\}\right]\nonumber\\
			&\qquad\qquad\qquad\quad\times e^{-\sum_{k}\int \dd t\int \dd\omega [b_k^*(\omega)Ab_k(\omega)+c_k^*(\omega)Ac_k(\omega)+i\xi_{b,k}b_k^*(\omega) o_k(t)+i\xi^*_{b,k}b_k(\omega) o_k^*(t)+i\xi_{c,k}c_k^*(\omega) o_k(t)+i\xi_{c,k}^*c_k(\omega) o_k^*(t)]}\nonumber\\
			&=Ne^{i\int \dd t\mathcal{L}_{\text{sys}}}e^{-\sum_{k}\int \dd t\int \dd t^\prime\int \dd\omega [\abs{\xi_{b,k}}^2o_k^*(t)\varPi_\omega(t-t^\prime)o_k(t^\prime)+\abs{\xi_{c,k}}^2o_k^*(t)\varPi_\omega(t-t^\prime)o_k(t^\prime)]}\nonumber\\
			&=Ne^{i\int \dd t\mathcal{L}_{\text{sys}}}e^{-\pi\sum_{k}\int \dd t  [\abs{\xi_{b,k}}^2o_k^*(t)o_k(t)+\abs{\xi_{c,k}}^2o_k^*(t)o_k(t)]}\label{Leff1},
		\end{align}
		where $N$ is the constant coefficient from the Gaussian functional integration. As a result, we can obtain the effective Lagrangian, and then the effective Hamiltonian is obtained from the effective Lagrangian by Legendre transformation, which is
		\begin{align}\label{Leff2}
			\mathcal{L}_{\text{eff}}=\mathcal{L}_{\text{sys}}+i\pi\sum_{k}(\abs{\xi_{b,k}}^2 +\abs{\xi_{c,k}}^2) o_k^*o_k\Longrightarrow H_{\text{eff}}=H_{\text{sys}}\{o_k\}-\frac{i}{2}\sum_k(\kappa_{b,k} +\kappa_{c,k})o_k^\dagger o_k.
		\end{align}
		Then by (\ref{Leff1}) and (\ref{Leff2}), Eq.~(\ref{path1}) can be simplified as 
		\begin{align}
			G^{\mu\nu}(t^\prime_{B_n};t_{D_n})
			&=(-1)^n\frac{\int\mathcal{D}\left[\left\{o_k,o_k^*\right\}\right]\prod_{l=1}^{n}\left[o_{\nu_l}(t^\prime_l)o^{*}_{\mu_l}(t_l)\right]e^{i\int \dd t\mathcal{L}_{\text{eff}}}}{\int\mathcal{D}\left[\left\{o_k,o_k^*\right\}\right]e^{i\int \dd t\mathcal{L}_{\text{eff}}}}\nonumber\\
			&=(-1)^n\frac{\int\mathcal{D}\left[\left\{o_k,o_k^*\right\}\right]\prod_{l=1}^{n}\left[\tilde{o}_{\nu_l}(t^\prime_l)\tilde{o}^{*}_{\mu_l}(t_l)\right]e^{i\int \dd t\mathcal{L}_{\text{eff}}}}{\int\mathcal{D}\left[\left\{o_k,o_k^*\right\}\right]e^{i\int \dd t\mathcal{L}_{\text{eff}}}}\nonumber\\
			&=(-1)^n\langle g| \mathcal{T}[\prod_{l=1}^{n}\tilde{o}_{\nu_l}(t^\prime_l)\tilde{o}^\dagger_{\mu_l}(t_l)]|g\rangle=\widetilde{G}^{\mu\nu}(t^\prime_{B_n};t_{D_n}),
		\end{align}
		where
		\begin{eqnarray}
			\begin{bmatrix}
				\tilde{o}_{\nu_l}(t)\\
				\tilde{o}^\dagger_{\mu_l}(t)
			\end{bmatrix}=\exp(iH_{\text{eff}}t)
			\begin{bmatrix}
				o_{\nu_l}\\
				o^\dagger_{\mu_l}
			\end{bmatrix}
			\exp(-iH_{\text{eff}}t).
		\end{eqnarray}
		Actually, we could use a more straightforward method to obtain the relation \cite{PhysRevA.95.063809}.
		
		\section{The derivation of probability amplitude with equal-time probing $n$ photons}\label{D}
		According to Eqs.~(\ref{eq11}) and (\ref{eq18}), the probability amplitude of equal-time probing $n$ outgoing photons can be written as 
		\begin{align}
			P_n^{\mu\nu}(t)&=\prod_{l=1}^{n}\int\frac{\dd{t^\prime_{l}}}{\sqrt{2\pi}}e^{-ik_{l}t^\prime_{l}}S^{\mu\nu}_{t\ldots t;t^\prime_1\ldots t^\prime_n}\nonumber\\
			&=\sum_{m=0}^n\sum_{B_m,D_m}\prod_{s=1}^{m}\int\frac{\dd{t^\prime_{D_m(s)}}}{\sqrt{2\pi}}e^{-ik_{D_m(s)}t^\prime_{D_m(s)}}G^{\mu_{D_m}\nu_{B_m}}(t_{B_m};t^\prime_{D_m})
			\sum_{P_c}\prod_{s=1}^{n-m}[\frac{e^{-ik_{P_cD^c_m(s)}t}}{\sqrt{2\pi}}\delta_{\nu_{B^c_m(s)},\mu_{P_cD^c_m(s)}}].\label{Pn}
		\end{align}
		
		In order to calculate Eq.~(\ref{Pn}), we first deal with the integral:
		\begin{align}
			I(D_n,B_n)\equiv\prod_{s=1}^{n}\int\frac{\dd{t^\prime_{D_n(s)}}}{\sqrt{2\pi}}e^{-ik_{D_n(s)}t^\prime_{D_n(s)}}G^{\mu_{D_n}\nu_{B_n}}(t_{B_n};t^\prime_{D_n}), \label{grenn_p}
		\end{align}
		where the time-ordered $2n$-point Green's function 
		\begin{align}
			G^{\mu_{D_n}\nu_{B_n}}(t_{B_n};t^\prime_{D_n})=G^{\mu\nu}(t\ldots t;t^\prime_1\ldots t^\prime_n)=\widetilde{G}^{\mu\nu}(t\ldots t;t^\prime_1\ldots t^\prime_n)=(-1)^n
			\langle g| \mathcal{T}[\prod_{j,k=1}^{n}{\tilde{o}_{\nu_{j}}(t)\tilde{o}^\dagger_{\mu_{k}}(t^\prime_{k})}]|g\rangle \label{grenn_t}.
		\end{align}
		Meanwhile, we can also take off the time-ordered operator in Eq.~(\ref{grenn_t}) but need to add all possible permutations, and thus the integral~(\ref{grenn_p}) is given by
		\begin{align}
			I(D_n,B_n)=(-1)^n\big[\prod_{s=1}^{n}\int\frac{\dd{t^\prime_s} }{\sqrt{2\pi}}e^{-ik_st^\prime_s}\big]\sum_{{\mathcal{P}}}\langle g|\prod_{j=1}^{n+1}\widetilde{\mathcal{O}}(t^\prime_{\mathcal{P}_j})|g\rangle\times\prod_{j=1}^{n}\theta(t^\prime_{\mathcal{P}_j}-t^\prime_{\mathcal{P}_{j+1}}),\label{gf}
		\end{align}
		where $\theta(x)$ is the step function, and $\mathcal{P}$ is permutations over indices $\{0,1,\ldots,n\}$. Note from Eq.~(\ref{gf}) that we have assumed $t_0^\prime=t$, $\widetilde{\mathcal{O}}(t^\prime_{0})=\prod_{j=1}^{n}\tilde{o}_{\nu_{j}}(t)$, and $\widetilde{\mathcal{O}}(t^\prime_{k})=\tilde{o}^\dagger_{\mu_{k}}(t^\prime_{k})$. According to Eqs.~(\ref{eq15}) and (\ref{eq17}), there are some permutations in Eq.~(\ref{gf}) that are automatically zero due to the property of upper and lower triangular matrices multiplication, e.g., $\tilde{o}_{\nu_m}(t_{\mathcal{P}_m})\tilde{o}_{\nu_n}(t_{\mathcal{P}_n})\tilde{o}^\dagger_{\mu_l}(t_{\mathcal{P}_l})|g\rangle=0$. In order to evaluate the integral about time, we need to introduce an integral trick and an identical equation:
		\begin{gather}
			\int_{-\infty}^{+\infty}e^{-iM(t-t^\prime)}e^{\pm ia(t-t^\prime)}\theta(t-t^\prime)\dd{t^\prime}=\lim\limits_{\varepsilon\to0^+}[i(M\mp a-i\varepsilon)]^{-1},\label{I1}\\
			\prod_{j=1}^{n}\int\frac{\dd{t_j}}{\sqrt{2\pi}}e^{-ik_jt_j}=\int \frac{\dd{t_{\widetilde{\mathcal{P}}_{1}}}}{\sqrt{2\pi}}e^{-it_{\widetilde{\mathcal{P}}_{1}}\sum_{s=1}^{n}k_{\widetilde{\mathcal{P}}_s}}\prod_{j=2}^{n}\frac{\dd{t_{\widetilde{\mathcal{P}}_j}}}{\sqrt{2\pi}}e^{i\sum_{s=j}^{n}k_{\widetilde{\mathcal{P}}_s}(t_{\widetilde{\mathcal{P}}_{j-1}}-t_{\widetilde{\mathcal{P}}_{j}})},\label{I2}
		\end{gather}
		where $M$ is square matrix, $a\in\mathbb{R}$, and $\widetilde{\mathcal{P}}$	are permutations over indices $\{1,2,\ldots,n\}$. Here, the imaginary part of all eigenvalues of $M$ is less than or equal to zero. Thereby, we can guarantee the validity of both $e^{-i(M-i\varepsilon)t}|_{t\to +\infty}=0$ and $\text{Det}[M\mp a-i\varepsilon]\neq0$. Based on Eqs.~(\ref{I1}) and (\ref{I2}), Eq.~(\ref{gf}) can be further simplified as
		\begin{align}
			&I(D_n,B_n)\nonumber\\
			&=(-1)^n\big[\prod_{s=1}^{n}\int\frac{\dd{t^\prime_s} }{\sqrt{2\pi}}e^{-ik_st^\prime_s}\big]\sum_{\widetilde{\mathcal{P}}}\langle g|\prod_{j=1}^{n}[e^{iH_{\text{eff}}t}o_{\nu_j}e^{-iH_{\text{eff}}t}]\prod_{j=1}^{n}[e^{iH_{\text{eff}}t^\prime_{\widetilde{\mathcal{P}}_j}}o^\dagger_{\mu_{\widetilde{\mathcal{P}}_j}}e^{-iH_{\text{eff}}t^\prime_{\widetilde{\mathcal{P}}_j}}]|g\rangle \theta(t-t^\prime_{\widetilde{\mathcal{P}}_{1}})\prod_{j=2}^{n}\theta(t^\prime_{\widetilde{\mathcal{P}}_{j-1}}-t^\prime_{\widetilde{\mathcal{P}}_{j}})\nonumber\\
			&=(-1)^n\sum_{\widetilde{\mathcal{P}}}\big[\prod_{j=1}^{n}\int\frac{\dd{t^\prime_{\widetilde{\mathcal{P}}_j}} }{\sqrt{2\pi}}\big]\langle g|\prod_{j=1}^{n}[o_{\nu_j}]e^{-iH_{\text{eff}}(t-t^\prime_{\widetilde{\mathcal{P}}_{1}})}e^{i(t-t_{\widetilde{\mathcal{P}}_{1}})\sum_{s=1}^{n}k_{\widetilde{\mathcal{P}}_s}}\theta(t-t^\prime_{\widetilde{\mathcal{P}}_{1}})e^{-it\sum_{s=1}^{n}k_{\widetilde{\mathcal{P}}_s}}\nonumber\\
			&\qquad \qquad\qquad\qquad\qquad\quad\ \ \times\prod_{j=1}^{n-1}[o^\dagger_{\mu_{\widetilde{\mathcal{P}}_{j}}}e^{-iH_{\text{eff}}(t^\prime_{\widetilde{\mathcal{P}}_{j}}-t^\prime_{\widetilde{\mathcal{P}}_{j+1}})}e^{i\sum_{s=j+1}^nk_{\widetilde{\mathcal{P}}_s}(t^\prime_{\widetilde{\mathcal{P}}_{j}}-t^\prime_{\widetilde{\mathcal{P}}_{j+1}})}\theta(t^\prime_{\widetilde{\mathcal{P}}_{j}}-t^\prime_{\widetilde{\mathcal{P}}_{j+1}})]o^\dagger_{\mu_{\widetilde{\mathcal{P}}_{n}}}|g\rangle \nonumber\\
			&=[\prod_{j=1}^{n}\frac{e^{-ik_jt}}{\sqrt{2\pi}}]\lim\limits_{\varepsilon\to0^+}\sum_{\widetilde{\mathcal{P}}}\langle g|\prod_{j=1}^{n}[o_{\nu_j}]\prod_{j=1}^{n}\{[-i(H_{\text{eff}}-\sum_{s=j}^{n}k_{\widetilde{\mathcal{P}}_s}-i\varepsilon)]^{-1}o^\dagger_{\mu_{\widetilde{\mathcal{P}}_j}}\}|g\rangle\nonumber\\
			&=[\prod_{j=1}^{n}\frac{e^{-ik_jt}}{\sqrt{2\pi}}]\sum_{\widetilde{\mathcal{P}}}[\overrightarrow{\prod}_{j=1}^{n}\textbf{O}^{\nu_{j}}_{j-1,j}][\overleftarrow{\prod}_{j={1}}^{n}\mathcal{K}_{\sum_{s=1}^{j}{k_{\widetilde{\mathcal{P}}_s}}}^{-1}(j)\textbf{O}^{\dagger \mu_{\widetilde{\mathcal{P}}_j}}_{j-1,j}],\label{gf1}
		\end{align}
		Analogously, the case of $D_m$ and $B_m$ also has
		\begin{align}
			I(D_m,B_m)=[\prod_{j=1}^{m}\frac{e^{-ik_{D_m(j)}t}}{\sqrt{2\pi}}]\sum_{P}[\overrightarrow{\prod}_{j=1}^{m}\textbf{O}^{\nu_{B_m(j)}}_{j-1,j}][\overleftarrow{\prod}_{j={1}}^{m}\mathcal{K}_{\sum_{s=1}^{j}{k_{PD_m(s)}}}^{-1}(j)\textbf{O}^{\dagger \mu_{PD_m(j)}}_{j-1,j}],\label{gf2}
		\end{align}
		
		In the first step above, we use Eq.~(\ref{eq15}). In the second step, we take advantage of $H_{\text{eff}}|g\rangle=0$ and Eq.~(\ref{I2}). In the third step, we use Eq.~(\ref{I1}). In the last step, we use Eq.~(\ref{eq16}). Meanwhile, the inverse matrix becomes well-defined due to the presence of the coefficient $\varepsilon$ when the effective Hamiltonian has one or more eigenvalues whose imaginary part is zero; in other words, $\text{Det}[H_{\text{eff}}-\omega-i\varepsilon]\neq0$. 
		
		Finally, let us plug Eq.~(\ref{gf2}) into Eq.~(\ref{Pn}), and we have 
		\begin{align}
			P_n^{\mu\nu}(t)&=\sum_{m=0}^n\sum_{D_m,B_m}I(D_m,B_m)\sum_{P_c}\prod_{s=1}^{n-m}[\frac{e^{-ik_{P_cD^c_m(s)}t}}{\sqrt{2\pi}}\delta_{\nu_{B^c_m(s)},\mu_{P_cD^c_m(s)}}]\nonumber\\
			&=\frac{e^{-ik_{\text{tot}}t}}{\sqrt{(2\pi)^n}}\sum_{m=0}^n\sum_{D_m,B_m}\sum_{P}[\overrightarrow{\prod}_{j=1}^{m}\textbf{O}^{\nu_{B_m(j)}}_{j-1,j}][\overleftarrow{\prod}_{j={1}}^{m}\mathcal{K}_{\sum_{s=1}^{j}{k_{PD_m(s)}}}^{-1}(j)\textbf{O}^{\dagger \mu_{PD_m(j)}}_{j-1,j}]\sum_{P_c}\prod_{s=1}^{n-m}[\delta_{\nu_{B^c_m(s)},\mu_{P_cD^c_m(s)}}].\label{Pn2}
		\end{align}
		
		For example, we assume $\mu=(b_i)^n$, $\nu=(c_j)^n$, and $k=(\omega_d)^n$. The corresponding input-output relations are written as $b_{i,\text{out}}(t)=b_{i,\text{in}}(t)-i\sqrt{\kappa_{b,i}}o_i(t)$ and $c_{j,\text{out}}(t)=c_{j,\text{in}}(t)-i\sqrt{\kappa_{c,j}}o_j(t)$. Obviously, the only non-zero term is for $m=n$, and we have
		\begin{align}
			P_n^{\mu\nu}(t)&=\left[\frac{e^{-i\omega_dt}}{\sqrt{2\pi}}\right]^n\sum_{D_n,B_n}\sum_{P}[\overrightarrow{\prod}_{l=1}^{n}\textbf{O}^{\nu_{B_n(l)}}_{l-1,l}][\overleftarrow{\prod}_{l={1}}^{n}\mathcal{K}_{\sum_{s=1}^{l}{k_{PD_n(s)}}}^{-1}(l)\textbf{O}^{\dagger \mu_{PD_n(l)}}_{l-1,l}]\nonumber\\
			&=n!\left[\frac{e^{-i\omega_dt}}{\sqrt{2\pi/(\kappa_{b,i}\kappa_{c,j})}}\right]^n\left[\overrightarrow{\prod}_{l=1}^{n}\textbf{O}_{l-1,l}^j\right]\left[\overleftarrow{\prod}_{l={1}}^{n}\mathcal{K}_{l\omega_d}^{-1}(l)\textbf{O}^{\dagger i}_{l-1,l}\right]. \label{Pn3}
		\end{align}
		Therefore, combining with the Eq.~(\ref{eq6}) and Eq.~(\ref{Pn3}), the $n$th-order equal-time correlation function under a weak coherent drive could be written as
		\begin{align}
			g^{(n)}_{jj}(0)&=\lim\limits_{\abs{\beta_i}\to0}\frac{\langle\psi_{\text{out}}|c_j^{\dagger n}(t)  c_j^n(t)|\psi_{\text{out}}\rangle}{\ \ \langle\psi_{\text{out}}| c_j^\dagger(t)c_j(t)|\psi_{\text{out}}\rangle ^n}=\lim\limits_{\abs{\beta_i}\to0}\frac{|\mathscr{N}|^2\sum_{k=n}^{\infty}\frac{|\beta_i|^{2k}}{k!}\ _{\omega_d}\langle \Psi_{\text{out}}^{(k)}|c_j^{\dagger n}(t)c_j^n(t)|\Psi_{\text{out}}^{(k)}\rangle_{\omega_d}}{\left[|\mathscr{N}|^2\sum_{k=1}^{\infty}\frac{|\beta_i|^{2k}}{k!}\ _{\omega_d}\langle \Psi_{\text{out}}^{(k)}|c_j^{\dagger}(t)c_j(t)|\Psi_{\text{out}}^{(k)}\rangle_{\omega_d} \right]^n}\nonumber\\
			&=\frac{_{\omega_d}\langle \Psi_{\text{out}}^{(n)}|c_j^{\dagger n}(t)c_j^n(t)|\Psi_{\text{out}}^{(n)}\rangle_{\omega_d}/n!}{\left[_{\omega_d}\langle \Psi_{\text{out}}^{(1)}|c_j^{\dagger}(t)c_j(t)|\Psi_{\text{out}}^{(1)}\rangle_{\omega_d} \right]^n}=\frac{|P_n^{\mu\nu}(t)/n!|^2}{|P_1^{\mu\nu}(t)|^{2n}}=\lim\limits_{\abs{\Omega_i}\to0}\frac{\text{Tr}[o_j^{\dagger n}o_j^n\rho_{ss}]}{\text{Tr}[o_j^\dagger o_j^{}\rho_{ss}]^n},
		\end{align}
		where $\beta_i=\Omega_i\sqrt{2\pi/\kappa_b}$, and $|\Psi_{\text{out}}^{(k)}\rangle_{\omega_d}=S|\Psi_{\text{in}}^{(k)}\rangle_{\omega_d}^{b_i}$. In the step above, we take advantage of the relation, i.e., $|\mathscr{N}|^2\to1$ when $|\beta_i|\to0$.
	
		\section{The discussions about the case of the nonlinear interaction between the system and the heat baths}\label{E}
				On the basis of Eq.~(\ref{eq1}), we assume that each local system also interacts nonlinearly with a new individual heat bath. Thus, the noninteraction Hamiltonian $H_{\text{B}}$ and interaction Hamiltonian $H_{\text{I}}$ should be rewritten as
				\begin{align}
					H_{\text{B}}&=\int \dd\omega \sum_k\omega\left[b_k^\dagger(\omega) b_k(\omega)+c_k^\dagger(\omega) c_k(\omega)+d_k^\dagger(\omega) d_k(\omega)\right],\\
					H_{\text{I}}&=\int\dd\omega\sum_k\left[\xi_{b,k}b_k^\dagger(\omega)o_k+\xi_{c,k}c_k^\dagger(\omega) o_k+\xi_{d,k}d_k^\dagger(\omega) o^m_k+\text{H.c.}\right],\label{E2}
				\end{align}
				where $d_k(\omega)$ ($d_k^\dagger(\omega)$) is bosonic annihilation (creation) operator of the new heat bath mode, and $m$ is a positive integer greater than 1. Notice that the total Hamiltonian $H_{\text{tot}}$ also respects the U(1) symmetry. Similarly, following Eqs.~(\ref{motion})–(\ref{A8}), one can develop the standard input-output formalism that relates $d_{l,\text{in}}$, $d_{l,\text{out}}$, and $o_l$ as
				\begin{align}
					d_{l,\text{out}}(t) = d_{l,\text{in}}(t)-i\sqrt{\kappa_{d,l}}o^m_l(t),
				\end{align} 
				where $\kappa_{d,l}=2\pi\abs{\xi_{d,l}}^2$. Subsequently, following Appendix \ref{C}, we could derive a new effective Hamiltonian, which is given by
				\begin{align}\label{E4}
					H_{\text{eff}}=H_{\text{sys}}\{o_k\}-\frac{i}{2}\sum_k(\kappa_{b,k}+\kappa_{c,k})o_k^\dagger o_k-\frac{i}{2}\sum_k\kappa_{d,k}o_k^{\dagger m}o_k^m.
				\end{align}
				Obviously, the nonlinear effect does not disappear, it just moved from the interaction Hamiltonian (\ref{E2}) to the effective Hamiltonian (\ref{E4}). Besides, if we still regard Eq.~(\ref{eq5}) as the initial state of the total Hamiltonian, according to Eqs.~(\ref{numerator})–(\ref{master_equation}), we will obtain a Lindblad master equation after tracing over the heat bath degrees of freedom, i.e.,
				\begin{align}
					\frac{\dd\rho_s}{\dd t}=-i[H_{\text{sys}}+H_{\text{d}}, \rho_s]+\sum_k(\kappa_{b,k}+\kappa_{c,k})\mathcal{D}[o_k]\rho_s+\sum_k\kappa_{d,k}\mathcal{D}[o_k^m]\rho_s,
				\end{align}
				where $H_\text{d}=[\Omega_i^*o_i\exp(i\omega_dt)+\text{H.c.}]$. Actually, we can also achieve a multi-photon driving case to a local system by changing the initial state of the total system, and the corresponding Hamiltonian and initial state are given by
				\begin{align}\label{E6}
					H_{\text{d}}=\mathcal{E}^*_lo_l^me^{i\omega_dt}+\mathcal{E}_lo_l^{\dagger m}e^{-i\omega_dt},\ |\psi_{\text{in}}\rangle=|\alpha_l\rangle_{\omega_d}^{d_l}\otimes|0\rangle_{\text{B}}\otimes|g\rangle,
				\end{align}
				where $\alpha_l=\mathcal{E}_l\sqrt{2\pi/\kappa_{d,l}}$, and $|0\rangle_\text{B}$ represents the vacuum state of baths except the mode $d_l$. 
				
				For the single-photon driving case [i.e., Eq.~(\ref{eq3})], our conclusions in the main text still apply for the nonlinear interaction, and we just need to replace the effective Hamiltonian with Eq.~(\ref{E4}). However, for the multi-photon driving case [i.e., Eq.~(\ref{E6})], the $n$th-order ETCF~(\ref{eq6}) under a weak drive is given by
				\begin{align}
					\tilde{g}_{jj}^{(n)}(0)&=\lim\limits_{\abs{\alpha_l}\to0}g_{jj}^{(n)}(0)=\lim\limits_{\abs{\alpha_l}\to0}\frac{|\mathscr{N}|^2\sum_{k=\left\lceil n/m\right\rceil}^{\infty}\frac{|\alpha_l|^{2k}}{k!}\ _{\omega_d}\langle \Psi_{\text{out}}^{(k)}|c_j^{\dagger n}(t)c_j^n(t)|\Psi_{\text{out}}^{(k)}\rangle_{\omega_d}}{\left[|\mathscr{N}|^2\sum_{k=1}^{\infty}\frac{|\alpha_l|^{2k}}{k!}\ _{\omega_d}\langle \Psi_{\text{out}}^{(k)}|c_j^{\dagger}(t)c_j(t)|\Psi_{\text{out}}^{(k)}\rangle_{\omega_d} \right]^n}\nonumber\\
					&=\mathbb{G}\times\lim\limits_{\abs{\alpha_l}\to0}|\alpha_l|^{-2(n-\left\lceil n/m\right\rceil)}=\mathbb{G}\times\infty=\infty,\label{E7}
				\end{align}
				where $\mathbb{G}$ is a constant that is entirely unrelated to $\alpha_l$, and $n-\left\lceil n/m\right\rceil$ is greater than 0 due to $m>1$. Thus, we must simultaneously add at least one single-photon drive term for the local system and select an appropriate parameters condition to prevent the correlation function from approaching infinity. To prove this point, let us introduce a driving term
				\begin{align}
					H_{\text{d}}=\Omega_i^* o_i e^{i\omega_dt}+\Omega_i o^\dagger_i e^{-i\omega_dt}+\mathcal{E}^*_lo_l^2 e^{2i\omega_d t}+\mathcal{E}_lo_l^{\dagger 2} e^{-2i\omega_d t},
				\end{align}
				and the corresponding input state is given by
				\begin{align}
					|\psi_{\text{in}}\rangle=|\beta_i\rangle_{\omega_d}^{b_i}\otimes|\alpha_l\rangle_{2\omega_d}^{d_l}\otimes|0\rangle_{\text{B}}\otimes|g\rangle.
				\end{align}
				For the sake of simplicity, we only provide the analytical expression of the second-order ETCF. To avoid the presence of the case of Eq.~(\ref{E7}), we must choose this parameter condition, i.e., $\alpha_l=\eta\beta_i^2$ with $\abs{\beta_i}\to 0$. Thus, we have
				\begin{align}
					\tilde{g}_{jj}^{(2)}(0)&=\lim\limits_{\abs{\beta_i}\to0}g_{jj}^{(2)}(0)=\frac{
						|\langle 0|[c_j(t)c_j(t)]S[b_i^\dagger(\omega_d)b_i^\dagger(\omega_d)]|0\rangle/2!+\eta\langle 0|[c_j(t)c_j(t)]S[d_l^\dagger(2\omega_d)]|0\rangle|^2
					}{|\langle 0|[c_j(t)]S[b_i^\dagger(\omega_d)]|0\rangle |^4}.\label{E10}
				\end{align}
				Notably, the denominator and the first term in the numerator can be analytically solved based on Eq.~(\ref{eq19}), and the corresponding expressions are given by
				\begin{align}
					\langle 0|[c_j(t)]S[b_i^\dagger(\omega_d)]|0\rangle&=\xi\times\textbf{O}^j_{0,1}\mathcal{K}_{\omega_d}^{-1}(1)\textbf{O}^{\dagger i}_{0,1},\label{E11}\\
					\langle 0|[c_j(t)c_j(t)]S[b_i^\dagger(\omega_d)b_i^\dagger(\omega_d)]|0\rangle&=2!\xi^2\times\textbf{O}^j_{0,1}\textbf{O}^j_{1,2}\mathcal{K}_{2\omega_d}^{-1}(2)\textbf{O}^{\dagger i}_{1,2}\mathcal{K}_{\omega_d}^{-1}(1)\textbf{O}^{\dagger i}_{0,1},\label{E12}
				\end{align}
				where $\xi=\sqrt{\kappa_{b,i}\kappa_{c,j}/2\pi}\exp({-i\omega_d t})$. For the second term in the numerator, we have
				\begin{align}
					\langle 0|[c_j(t)c_j(t)]S[d_l^\dagger(2\omega_d)]|0\rangle&=\int\frac{\dd{t^\prime}}{\sqrt{2\pi}}e^{-2i\omega_dt^\prime}\langle0|c_{j,\text{out}}(t)c_{j,\text{out}}(t)d^\dagger_{l,\text{in}}(t^\prime)|0\rangle\nonumber\\
					&=\int\frac{\dd{t^\prime}}{\sqrt{2\pi}}e^{-2i\omega_dt^\prime}\left(i\kappa_{c,j}\sqrt{\kappa_{d,l}}\right)\langle g|\mathcal{T}[\tilde{o}_{i}(t)\tilde{o}_{i}(t)\tilde{o}^\dagger_{l}(t^\prime)\tilde{o}_{l}^\dagger(t^\prime)]|g\rangle\nonumber\\
					&=\int\frac{\dd{t^\prime}}{\sqrt{2\pi}}e^{-2i\omega_dt^\prime}\left(i\kappa_{c,j}\sqrt{\kappa_{d,l}}\right)\langle g|[o_{i}o_{i}]e^{-iH_{\text{eff}}(t-t^\prime)}[o_{l}^\dagger o_{l}^\dagger]|g\rangle\theta(t-t^\prime)\nonumber\\
					&=e^{-2i\omega_dt}\left(-i\kappa_{c,j}\sqrt{\kappa_{d,l}/2\pi}\right)\mathbf{O}^j_{0,1}\mathbf{O}^j_{1,2}\mathcal{K}_{2\omega_d}^{-1}(2)\mathbf{O}^{\dagger l}_{1,2}\mathbf{O}^{\dagger l}_{0,1}.\label{E13}
				\end{align}
				Finally, let us plug Eqs.~(\ref{E11})–(\ref{E13}) into (\ref{E10}), and the second-order ETCF can be simplified as
				\begin{align}
					\tilde{g}_{jj}^{(2)}(0)=\frac{\big|\textbf{O}^j_{0,1}\textbf{O}^j_{1,2}\mathcal{K}_{2\omega_d}^{-1}(2)\textbf{O}^{\dagger i}_{1,2}\mathcal{K}_{\omega_d}^{-1}(1)\textbf{O}^{\dagger i}_{0,1}-i\tilde{\eta}\times\mathbf{O}^j_{0,1}\mathbf{O}^j_{1,2}\mathcal{K}_{2\omega_d}^{-1}(2)\mathbf{O}^{\dagger l}_{1,2}\mathbf{O}^{\dagger l}_{0,1}\big|^2}{\big|\textbf{O}^j_{0,1}\mathcal{K}_{\omega_d}^{-1}(1)\textbf{O}^{\dagger i}_{0,1}\big|^4},
				\end{align}
				where $\tilde{\eta}=\eta\sqrt{2\pi\kappa_{d,l}}/\kappa_{b,i}=\mathcal{E}_l/\Omega_i^2$.
				
	\end{widetext}
	%

\end{document}